\documentclass[twocolumn]{aastex631}

\shorttitle{XMPGs in DESI}
\shortauthors{H. Zou et al.}
\graphicspath{{./}{figures/}}
\usepackage{amsmath}
\usepackage{autobreak}
\usepackage{makecell}
\usepackage{graphicx}
\newcommand{\Ha}{\mbox{H$\alpha$}}      
\newcommand{\Hb}{\mbox{H$\beta$}}       
\newcommand{\Hg}{\mbox{H$\gamma$}}      
\newcommand{\HII}{\mbox{\ion{H}{2}}}    
\newcommand{\SII}{[\mbox{\ion{S}{2}}]}    
\newcommand{\OIII}{[\mbox{\ion{O}{3}}]}   
\newcommand{\NII}{[\mbox{\ion{N}{2}}]}   
\newcommand{\NIIS}{[\mbox{\ion{N}{2}}]$\lambda$6583}
\newcommand{\OIIIFIZ}{[\mbox{\ion{O}{3}}]$\lambda$5007}
\newcommand{\OIIIFON}{[\mbox{\ion{O}{3}}]$\lambda$4959}
\newcommand{\OIIIFOT}{[\mbox{\ion{O}{3}}]$\lambda$4363}
\newcommand{\OIITS}{[\mbox{\ion{O}{2}}]$\lambda$3726}
\newcommand{\OIITN}{[\mbox{\ion{O}{2}}]$\lambda$3729}

\newcommand{\OII}{[\mbox{\ion{O}{2}}]}    

\newcommand{\Te}{\mbox{$T_{\rm e}$}}
\newcommand{\Ne}{\mbox{$N_{\rm e}$}}

\newcommand{\Msun}{\mbox{$M_{\odot}$}}

\newcommand{\Zsun}{\mbox{$Z_{\odot}$}}

\newcommand{\boldtext}[1]{\textcolor[rgb]{0,0,0}{#1}}

\begin{document}

\title{A Large Sample of Extremely Metal-poor Galaxies at $z<1$ Identified from the DESI Early Data}

\author[0000-0002-6684-3997]{Hu Zou}
\affiliation{Key Laboratory of Optical Astronomy, National Astronomical Observatories, Chinese Academy of Sciences, Beijing 100012, China}
\affiliation{School of Astronomy and Space Science, University of Chinese Academy of Sciences, Beijing 101408, China}
\author{Jipeng Sui}
\affiliation{Key Laboratory of Optical Astronomy, National Astronomical Observatories, Chinese Academy of Sciences, Beijing 100012, China}
\affiliation{School of Astronomy and Space Science, University of Chinese Academy of Sciences, Beijing 101408, China}
\author[0000-0003-4357-3450]{Am\'elie Saintonge}
\affiliation{Department of Physics \& Astronomy, University College London, Gower Street, London, WC1E 6BT, UK}
\author[0000-0002-6867-1244]{Dirk Scholte}
\affiliation{Department of Physics \& Astronomy, University College London, Gower Street, London, WC1E 6BT, UK}
\author[0000-0002-2733-4559]{John Moustakas}
\affiliation{Department of Physics and Astronomy, Siena College, 515 Loudon Road, Loudonville, NY 12211, USA}
\author{Malgorzata Siudek}
\affiliation{Institute of Space Sciences, ICE-CSIC, Campus UAB, Carrer de Can Magrans s/n, 08913 Bellaterra, Barcelona, Spain}
\author{Arjun Dey}
\affiliation{NSF's NOIRLab, 950 N. Cherry Ave., Tucson, AZ 85719, USA}
\author{Stephanie Juneau}
\affiliation{NSF's NOIRLab, 950 N. Cherry Ave., Tucson, AZ 85719, USA}
\author[0000-0001-9457-0589]{Weijian Guo}
\affiliation{Key Laboratory of Optical Astronomy, National Astronomical Observatories, Chinese Academy of Sciences, Beijing 100012, China}
\author{Rebecca Canning}
\affiliation{Institute of Cosmology \& Gravitation, University of Portsmouth, Dennis Sciama Building, Portsmouth, PO1 3FX, UK}
\author{J.~Aguilar}
\affiliation{Lawrence Berkeley National Laboratory, 1 Cyclotron Road, Berkeley, CA 94720, USA}
\author[0000-0001-6098-7247]{S.~Ahlen}
\affiliation{Physics Dept., Boston University, 590 Commonwealth Avenue, Boston, MA 02215, USA}
\author{D.~Brooks}
\affiliation{Department of Physics \& Astronomy, University College London, Gower Street, London, WC1E 6BT, UK}
\author{T.~Claybaugh}
\affiliation{Lawrence Berkeley National Laboratory, 1 Cyclotron Road, Berkeley, CA 94720, USA}
\author{K.~Dawson}
\affiliation{Department of Physics and Astronomy, The University of Utah, 115 South 1400 East, Salt Lake City, UT 84112, USA}
\author[0000-0002-1769-1640]{A.~de la Macorra}
\affiliation{Instituto de F\'{\i}sica, Universidad Nacional Aut\'{o}noma de M\'{e}xico, Cd. de M\'{e}xico C.P. 04510, M\'{e}xico}
\author{P.~Doel}
\affiliation{Department of Physics \& Astronomy, University College London, Gower Street, London, WC1E 6BT, UK}
\author[0000-0002-2890-3725]{J.~E.~Forero-Romero}
\affiliation{Observatorio Astron\'omico, Universidad de los Andes, Cra. 1 No. 18A-10, Edificio H, CP 111711 Bogot\'a, Colombia}
\author[0000-0003-3142-233X]{S.~Gontcho A Gontcho}
\affiliation{Lawrence Berkeley National Laboratory, 1 Cyclotron Road, Berkeley, CA 94720, USA}
\author{K.~Honscheid}
\affiliation{Department of Physics, The Ohio State University, 191 West Woodruff Avenue, Columbus, OH 43210, USA}
\author[0000-0003-1838-8528]{M.~Landriau} 
\affiliation{Lawrence Berkeley National Laboratory, 1 Cyclotron Road, Berkeley, CA 94720, USA}
\author[0000-0001-7178-8868]{L.~Le~Guillou}
\affiliation{Sorbonne Universit\'{e}, CNRS/IN2P3, Laboratoire de Physique Nucl\'{e}aire et de Hautes Energies (LPNHE), FR-75005 Paris, France}
\author[0000-0003-4962-8934]{M.~Manera}
\affiliation{Institut de F\'{i}sica d’Altes Energies (IFAE), The Barcelona Institute of Science and Technology, Campus UAB, 08193 Bellaterra Barcelona, Spain}
\author[0000-0002-1125-7384]{A.~Meisner}
\affiliation{NSF's NOIRLab, 950 N. Cherry Ave., Tucson, AZ 85719, USA}
\author{R.~Miquel}
\affiliation{Institut de F\'{i}sica d’Altes Energies (IFAE), The Barcelona Institute of Science and Technology, Campus UAB, 08193 Bellaterra Barcelona, Spain}
\author[0000-0001-6590-8122]{Jundan Nie}
\affiliation{Key Laboratory of Optical Astronomy, National Astronomical Observatories, Chinese Academy of Sciences, Beijing 100012, China}
\author{C.~Poppett}
\affiliation{Space Sciences Laboratory, University of California, Berkeley, 7 Gauss Way, Berkeley, CA 94720, USA}
\author[0000-0001-5589-7116]{M.~Rezaie}
\affiliation{Department of Physics, Kansas State University, 116 Cardwell Hall, Manhattan, KS 66506, USA}
\author{G.~Rossi}
\affiliation{Department of Physics and Astronomy, Sejong University, Seoul, 143-747, Korea}
\author[0000-0002-9646-8198]{E.~Sanchez}
\affiliation{Barcelona-Madrid RPG - Centro de Investigaciones Energéticas, Medioambientales y Tecnológicas}
\author{M.~Schubnell}
\affiliation{Department of Physics \& Astronomy, Ohio University, Athens, OH 45701, USA}
\author[0000-0002-6588-3508]{H.~Seo}
\affiliation{Department of Physics \& Astronomy, Ohio University, Athens, OH 45701, USA}
\author[0000-0003-1704-0781]{G.~Tarl\'{e}}
\affiliation{University of Michigan, Ann Arbor, MI 48109, USA}
\author[0000-0002-4135-0977]{Zhimin Zhou}
\affiliation{Key Laboratory of Optical Astronomy, National Astronomical Observatories, Chinese Academy of Sciences, Beijing 100012, China}
\author[0000-0002-3983-6484]{Siwei Zou}
\affiliation{Department of Astronomy, Tsinghua University, 30 Shuangqing Road, Haidian District, Beijing, China, 100190}



\begin{abstract}
Extremely metal-poor galaxies (XMPGs) at relatively low redshift are excellent laboratories for studying galaxy formation and evolution in the early universe. Much effort has been spent on identifying them from large-scale spectroscopic surveys or spectroscopic follow-up observations. Previous work has identified a few hundred XMPGs. In this work, we obtain a large sample of 223 XMPGs at $z<1$ from the early data of the Dark Energy Spectroscopic Instrument (DESI). The oxygen abundance is determined using the direct {\Te} method based on the detection of the {\OIIIFOT} line. The sample includes 95 confirmed XMPGs based on the oxygen abundance uncertainty; remaining 128 galaxies are regarded as XMPG candidates. \boldtext{These XMPGs are only 0.01\% of the total DESI observed galaxies. Their coordinates and other proprieties are provided in the paper.} The most XMPG has an oxygen abundance of $\sim 1/34\Zsun$, stellar mass of about $1.5\times10^7 \Msun$ and star formation rate of 0.22 {\Msun} yr$^{-1}$. The two most XMPGs present distinct morphologies suggesting different formation mechanisms. The local environmental investigation shows that XMPGs preferentially reside in relatively low-density regions. Many of them fall below the stellar mass-metallicity relations (MZRs) of normal star-forming galaxies. From a comparison of the MZR with theoretical simulations, it appears that XMPGs are good analogs to high-redshift star-forming galaxies. The nature of these XMPG populations will be further investigated in detail with larger and more complete samples from the on-going DESI survey.
\end{abstract}

\keywords{Metal abundances (1031) --- Scaling relations (2031) --- Dwarf galaxies (416) --- Galaxy surveys (1378)}


\section{Introduction} \label{sec:intro}
The hierarchical theory of galaxy formation suggests that large galaxies form through the assembly of smaller ones \citep{Whi91,Col00}. It is expected that there are a vast number of low-mass galaxies with low metal content in the early universe. Extremely metal-poor galaxies (XMPGs), which are chemically unevolved galaxies, provide ideal laboratories for validating the chemical evolution theories of galaxies and studying the physical processes in the early stages of their evolution. These galaxies are quite common at high redshift, but difficult to observe due to their low masses. The metallicities of galaxies at high redshift are also hard to reliably measure without large telescopes \citep[e.g.,][]{San16b,Gbu19}, with JWST also now making this possible \citep[e.g.][]{Cur23}. Thus, local metal-poor galaxies as possible analogs of primeval high-redshift young galaxies in terms of mass and metallicity, provide rich details for understanding the galaxy properties in extreme conditions, mass assembly and elemental abundance in the early universe. \boldtext{We should also note that high-redshift dwarfs are formed in extreme over-dense environment and the star formation process could be quite different in these conditions \citep[e.g.,][]{Dek23}.}  XMPGs have been used to explore chemical abundances, dust content, star formation, stellar feedback, escape of ionizing radiation, and cosmic environment in the low-metallicity conditions \citep{Thu99,Her12,Fil15,Sac16,Olm17,Pla19,Xuy22}. Because of the early stage of galactic chemical evolution, they have been used to determine the primordial He abundance, providing critical constraints on big bang nucleosynthesis \citep{Izo04,Hsy20,Mat22}. 

XMPGs are defined as galaxies with metallicity $Z < 0.1\Zsun$ \boldtext{(hereafter the metallitity is referred to the gas-phase metallicity in {\HII} regions).} The most reliable metallicity measurements come from the so-called $T_\mathrm{e}$ method, which relies on the detection of the {\OIIIFOT} auroral line. The line intensity ratio of {\OIIIFOT/\OIIIFIZ} is sensitive to the electron temperature and hence the \boldtext{estimated} oxygen abundance \citep{All84,Nic14}. XMPGs are extremely rare and they constitute only about 0.1\% of all galaxies in the local volume \citep{Mor11}.  Enormous efforts have been made during the past decades to identify XMPGs based on the detection of the {\OIIIFOT} line in optical spectra. However, the {\OIIIFOT} line is usually very weak, making the discovery of XMPGs difficult. Before the review of \citet{Kun00}, it was reported that only 31 galaxies reached the metallicity threshold of XMPGs. \citet{Mor11} identified a set of 130 XMPGs after a thorough bibliographic search and discovered 11 new ones with negligible {\NII} lines in the seventh data release of the Sloan Digital Sky Survey (SDSS). \boldtext{This is because the {\NII}/{\Ha} ratio is highly sensitive and positively correlate to the metallicity \citep{Den02}.} Using the same spectroscopic data set, \citet{San16a} has produced by far the largest sample of 196 XMPGs with intense {\OIIIFOT} with respect to {\OIIIFIZ}. A few more XMPGs have been discovered in later SDSS data releases \citep[e.g., ][]{Gao18} and other spectroscopic surveys, such as DEEP2 \citep{LyC15}, LAMOST \citep{Gao17} and VUDS \citep{Amo14}. In addition to directly detecting the {\OIIIFOT} line in the spectroscopic data, some authors attempted to select XMPG candidates from multi-wavelength photometric data and confirmed them by spectroscopic follow-ups \citep{Bro08,Hsy18,Koj20,Iso22}. Several tens of XMPGs are found in this way. There are a total of about 350 known XMPGs collected from literature by Sui et al. (in preparation), \boldtext{who will} utilize deep-learning algorithms to identify XMPG candidates from DESI Legacy Imaging Surveys \citep{Zou17,Dey19}.  

The number of existing XMPGs remains small. Systematically searching for XMPGs in large-scale spectroscopic surveys will considerably increase the sample size, improve the completeness, and provide an opportunity to take a systematic census of such an extreme galaxy population. DESI is embarking on a five-year spectroscopic survey to explore the nature of dark energy \citep{Lev13,Des16a,Des16b}. It is observing about 40 million extragalactic objects, including bright galaxies, luminous red galaxies (LRGs), emission line galaxies (ELGs), and quasars (QSOs) \citep{Des16a,Sch23,Mye23}. The DESI Bright Galaxy Survey (BGS) will produce a magnitude-limited galaxy sample down to $r<19.5$ mag \citep{Hah23}, helping us to obtain a uniform XMPG sample at low redshift, while the ELG survey \citep{Rai23b} can be used for detecting XPMGs at higher redshift. It is expected that several thousands XMPGs will be identified in the DESI survey, \boldtext{which is estimated from our XMPG detection with the DESI early data in this paper.} We will use the direct $T_\mathrm{e}$ method to measure the metallicities of galaxies to identify XMPGs at $z<1$. With this sample, we can further study their photometric and spectral properties, investigate their local and global environment and gas component, and explore the scaling relations, star formation activity and gas flows that are associated with such low-metallicity conditions. 

This work focuses on the analysis of DESI early data, including the Early Data Release \citep[EDR;][]{Des23b}, the detection of XMPGs and exploration of their mass-metallicity relation. The paper is organized as follows. In Section \ref{sec:data}, we introduce the DESI spectroscopic data and our galaxy sample selection. Section \ref{sec:measure} presents the determination of oxygen abundance and measurements of stellar mass and star formation rate. The XMGPs identified in this work and their basic properties are shown in Section \ref{sec:xmpgs}. The XMPG distribution on the plane of stellar mass-metallicity relation is also discussed in this section. The summary is given in Section \ref{sec:summary}. Throughout this paper, we assume a flat $\Lambda$-CDM cosmology with $\Omega_\Lambda =0.7$, $\Omega_m=0.3$, and $H_0$ = 70 km s$^{-1}$ Mpc$^{-1}$. Both photometric magnitudes and spectral line fluxes are corrected for the Galactic extinction. The solar Oxygen abundance (12+log(O/H)$_\odot$ = 8.69) adopted in this paper is from \citet{Asp21}.  

\section{Galaxy Sample Selection} \label{sec:data}
\subsection{DESI spectroscopic data}

The DESI experiment is a \boldtext{new generation ground-based} cosmological survey, aiming to explore the expansion history and structure growth rate of the universe with unprecedented precision and to understand the nature of dark energy \citep{Lev13,Des16a,Des16b,Des23a,Des23b}.  It will accurately measure the redshifts of about 40 million galaxies and quasars over a 14,000 square degree footprint. The instrument is a multi-object spectroscopic system installed at the prime focus of the Mayall 4m telescope at Kitt Peak, Arizona \citep{Des22,Sil23,Mil23}. It features a 3.2-deg field-of-view and 5000 fiber robots that can position optical fibers to 5000 objects and collect their spectra simultaneously. The fibers (diameter of 1.5\arcsec) are fed to ten 3-arm spectrographs, which provide a continuous wavelength coverage of 3600--9800\AA.  The spectral resolution varies from $R\sim$~2000 in the blue end to 5000 in the red, which is high enough to resolve the {\OII} doublet. 
 
The DESI spectroscopic targets include approximately 50 million extragalactic and Galactic objects, mainly selected from the DESI Legacy Imaging Surveys \citep{Zou17, Dey19, Sch23}. The imaging data are about 2 mag deeper than the SDSS, covering a sky area of 20,000 deg$^2$ in optical $grz$ bands and including the latest six-year near-infrared observations from the Wide-field Infrared Survey Explorer (WISE) satellite. The primary targets to be observed in dark time are LRGs at $0.4<z<1.1$, ELGs at $0.6<z<1.6$ and QSOs at $z<3.5$ \citep{Rai20,Rui20,Yec20,Zho20,Cha23,Zho23,Rai23b}. During bright time, DESI will carry the Bright Galaxy Survey (BGS) of galaxies at $z<0.6$ and Milky Way survey of stars \citep{All20,Rui20,Coo22,Hah23}. There are also some secondary target programs to utilize the spare fibers and DESI capabilities, such as the LOW-Z Secondary Target Survey targeting low-redshift ($z<0.03$) dwarf galaxies \citep{Dar22}.

\begin{figure*}[ht!]
\centering
\includegraphics[width=0.9\textwidth]{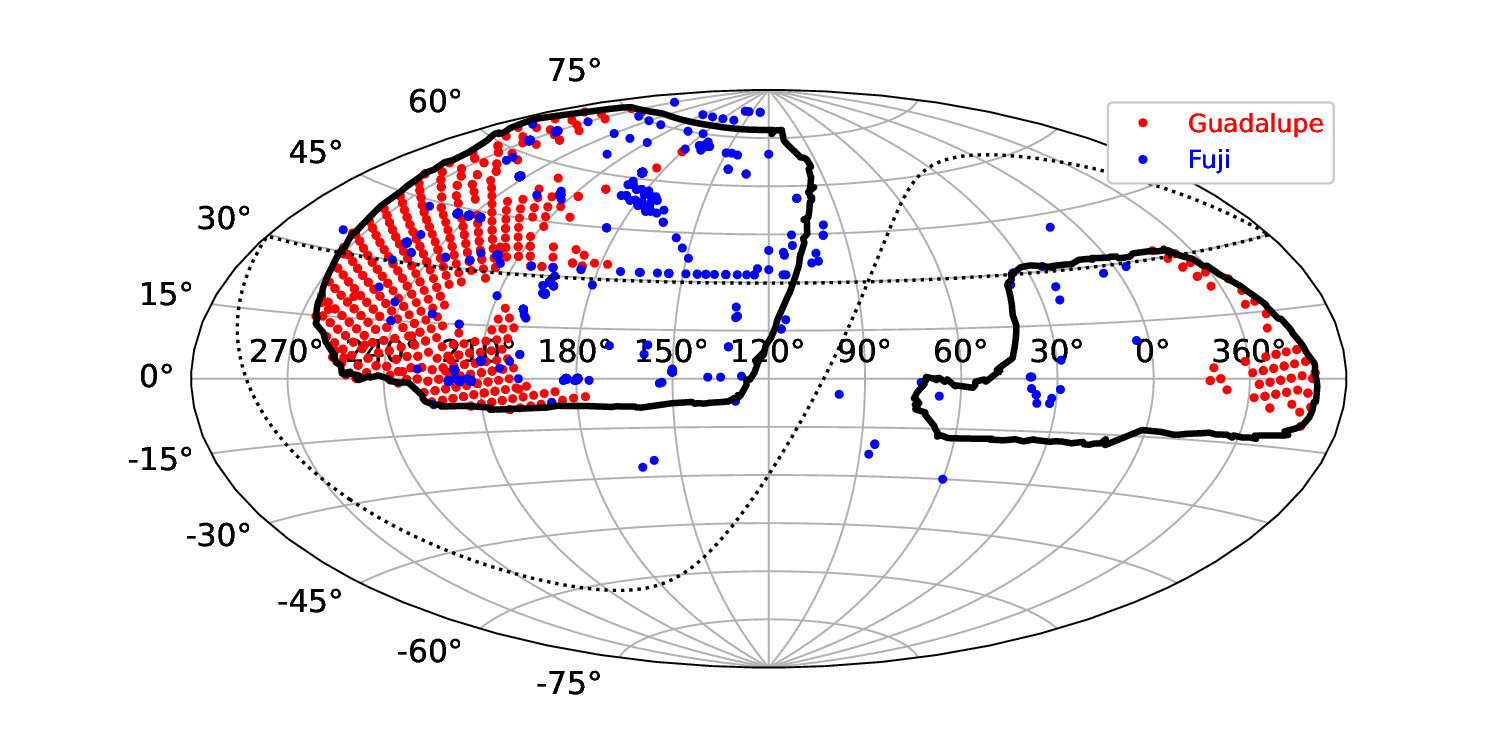}
\caption{Sky coverage of the DESI early data in Aitoff projection. The black lines delineate the DESI baseline footprint of about 14,000 square degrees. The dotted line shows the Galactic plane. The circles represent the centers of 3.2-degree DESI tiles (blue for \textit{Fuji} and red for \textit{Guadalupe}). Note that a few \textit{Fuji} tiles are located outside of the DESI footprint. \label{fig:coverage}}
\end{figure*}

The DESI project is expected to be a five year survey. Before the main survey started on May 14, 2021, there were several stages of survey validation for about six months \citep{Des23a}, which aimed to refine the target selections and fiber assignment, improve the data reduction pipeline and observing efficiency, and validate the instrumental performances \citep{Ale23,Bai23,Des23a,Kir23,Rai23a,Slf23,Lan23}, etc. An automatic spectroscopic data pipeline has been developed to reduce raw observations into wavelength- and flux-calibrated spectra, and to perform redshift measurements for all observed targets \citep{Guy23}.  The spectroscopic classifications and redshifts are determined using the \textit{Redrock} pipeline\footnote{\url{https://github.com/desihub/redrock}} \citep{Bai23,Bro23}, which fits a suite of templates of stars, galaxies, and quasars to the DESI spectra. There are several internal data releases generated by these data reduction and analysis pipelines, such as \textit{Fuji} data assembly for the survey validation (732 observed tiles) and \textit{Guadalupe} data assembly for the first two-month of the main surveys (653 observed tiles). Although the Early Data Release (EDR) only includes the \textit{Fuji} data \citet{Des23b}, we consider both \textit{Fuji} and \textit{Guadalupe} data assemblies as the DESI early data in this paper. Figure \ref{fig:coverage} shows the distribution of the observed DESI tiles in the full DESI footprint. 

\subsection{Flux measurements of emission lines}
The underlying stellar continuum is reproduced by using a full spectral fitting code of STARLIGHT \citet{Cid05}, where the combination of 45 single stellar populations (SSPs) from \citet{Bru03} and the \citet{Cha03} initial mass function (IMF) are adopted. The SSPs include 15 ages ranging from 1 Myr to 13 Gyr and three different metallicities (0.01, 0.02, 0.05). During the fitting, the main optical emission lines are masked out. 
 
\begin{figure*}[ht!]
\centering
\includegraphics[width=0.9\textwidth]{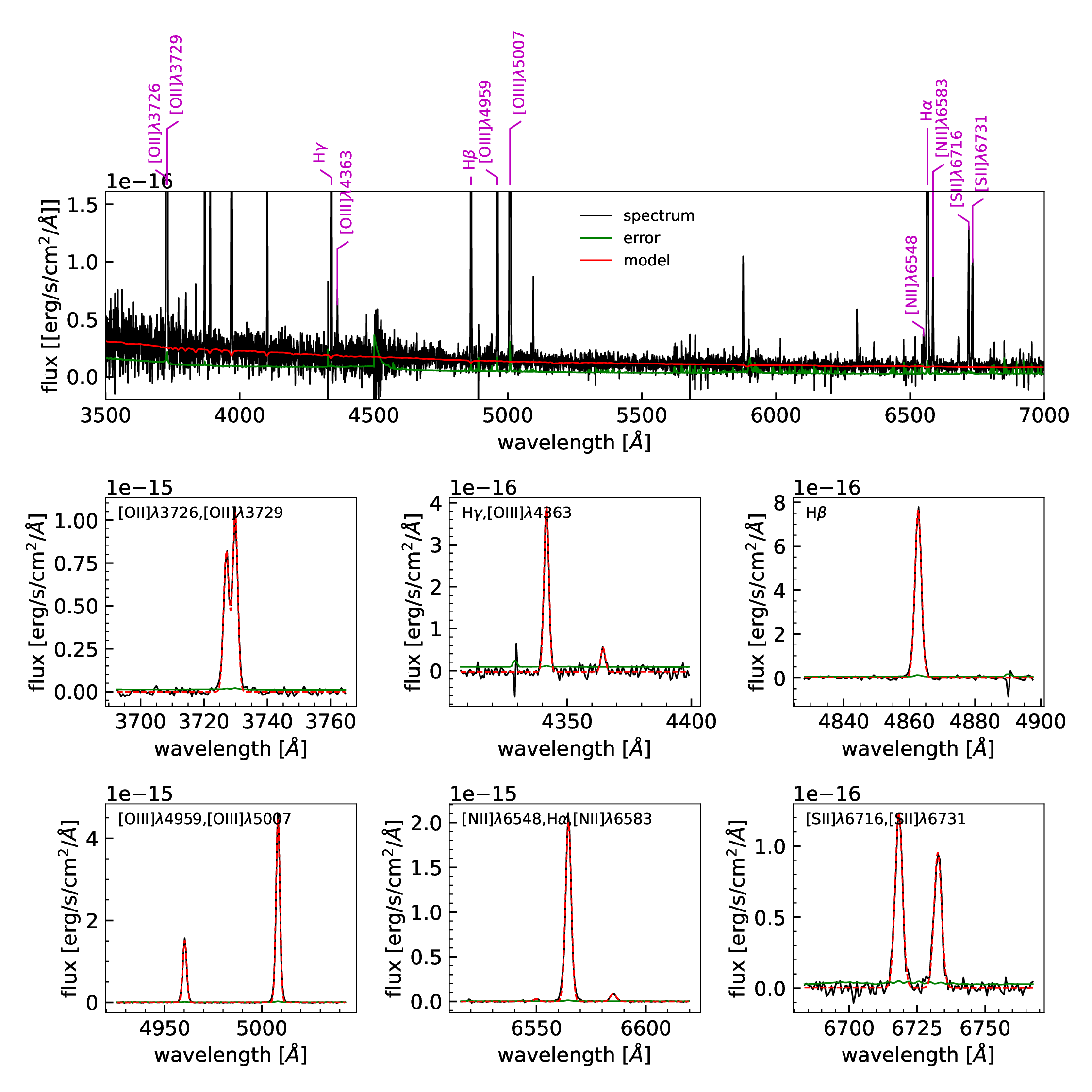}
\caption{Example DESI galaxy spectrum and demonstration of the emission line flux measurement technique. The top panel shows the spectrum of an arbitrarily-selected galaxy at $z=0.29$. The black line is the observed spectrum in rest frame, the green line is the corresponding error spectrum, and the red line is the best-fitted model spectrum. The middle and bottom rows display local parts of the residual spectrum (observed - model) around a specified line or group of lines. Each line is fitted with a Gaussian function, and the grouped lines are fitted with multi-Gaussian functions simultaneously. The red dashed curve in each panel is the best-fitted line profile, which is a combination of Gaussian function and a linear background.   \label{fig:spec}}
\end{figure*}

The modelled stellar continuum is subtracted from the observed spectrum and the fluxes of emission lines are then measured in the residual spectrum by Gaussian profile fitting. If not explicitly specified, all the spectral line measurements are calculated in rest frame. We are interested in the following emission lines: {\OII${\lambda\lambda}$3727,3729}, {\Hg}, {\OIIIFOT}, {\Hb}, {\OIII$\lambda\lambda$4959,5007}, {\NII${\lambda\lambda}$6548,6583}, {\Ha}, and {\SII${\lambda\lambda}$6716,6731}. Due to their proximity, the following line groups are defined: {\OII${\lambda\lambda}$3727,3729}; {\Hg} and {\OIIIFOT}; {\OIII$\lambda\lambda$4959,5007}; {\NII${\lambda\lambda}$6548,6583} and {\Ha}; {\SII${\lambda\lambda}$6716,6731}. Emission line fluxes in each group are fitted with Gaussian functions simultaneously in order to mitigate the contamination of neighboring lines. The steps of emission-line measurements are described below: %
\begin{enumerate}
\item The local background is estimated by a linear fit to the spectrum within a 70{\AA} wavelength window around each line or line group, excluding regions affected by emission lines (15{\AA} width). The linear fit is performed with a robust 3$\sigma$ clipping algorithm. 
\item The fitted local linear background is subtracted from the spectrum and the background error is estimated from the RMS of the linear fit.  
\item The background-subtracted emission line profiles (within a 15{\AA} window) are modeled with Gaussian functions, where the mean, standard deviation and amplitude are free parameters to be fitted. The parameter uncertainties are provided by Gaussian fitting with the error spectrum as the weight. 
\item In addition to the total flux, Gaussian width, and line centers, we also calculate the rest-frame equivalent width for each emission line, whose error is estimated through error propagation.
\end{enumerate}
Figure \ref{fig:spec} shows \boldtext{a randomly selected DESI galaxy spectrum} in rest-frame wavelength and the Gaussian fits to the emission-line profiles. 

\subsection{{\OIIIFOT}-detected Galaxy Sample}
There are a total of 2,856,369 unique galaxies with successful measurements of spectral lines in the \textit{Fuji} and \textit{Gaudalupe} data sets. These objects are selected from the redshift catalogs with SPECTYPE $=$ ``GALAXY'' and ZWARN $= 0$, which means that they are spectroscopically confirmed galaxies with reliable redshift measurements. Table \ref{tab:selection} presents the sample selection steps and criteria. \boldtext{In order to eliminate spurious emission line detections due to cosmic rays and large random noise, we only use emission lines with Gaussian FWHM $>$ 1.2\AA, which is empirically set to be slightly larger than the pixel scale of 0.8 {\AA} in observed wavelength.} To obtain credible metallicity measurements as shown in Section \ref{sec:measure}, we require that the {\OIIIFOT}, {\OIITS}, {\OIITN}, {\OIIIFON}, and {\OIIIFIZ} flux measurements have high-enough SNRs as shown in Table \ref{tab:selection}. Due to the lower limit of SNR(\OIIIFOT), the fitted FWHMs of this line might be abnormally large in some cases. In addition, the DESI data pipeline might mistakenly classify some broad-line AGNs as ``GALAXY". So we set an upper FWHM limit of 3 {\AA} for {\OIIIFOT} and \Hb, although it might exclude some galaxies with galactic outflow that would be worth further investigation in the future. The inclusion of {\OIIIFIZ} limits our galaxies to the redshift range of $z \lesssim 0.96$. We obtain a total of 1,633 {\OIIIFOT}-selected galaxies that satisfy all the above SNR and FWHM cuts.

\begin{deluxetable*}{ccc}
\centering
\tablecaption{Sample selection of \OIIIFOT-detected star-forming galaxies.\label{tab:selection}}
\tablehead{\colhead{Selection steps} & \colhead{Selection criteria} & \colhead{Number of galaxies}}
\startdata  
galaxies with reliable redshifts & \makecell{SPECTYPE $==$ GALAXY \\ ZWARN $==$ 0} & 2,856,369 \\
\tableline
SNR and FWHM cuts & \makecell{SNR(\OIIIFOT) $>3$ \\ SNR(\OIIIFON) $>3$ $\|$ SNR(\OIIIFIZ) $>3$ \\ SNR(\OIITS) $>3$ $\|$ SNR(\OIITN) $>3$ \\ SNR(\Hb) $>3$ \\ SNR(\Ha) $>3$ $\|$ SNR(\Hg) $>3$ \\ FWHM(\OIII, \OII, \Ha, \Hb, \Hg, \SII, \NII) $>1.2$ \AA \\ FWHM(\OIIIFOT, \Hb, \OIIIFIZ) $<3$ \AA } & 1,633 \\
\tableline
BPT diagnostic diagrams & \makecell{star-forming galaxies in BPT diagrams: \\ \OIII/{\Hb} vs. \NII/{\Ha}  \&\& \OIII/{\Hb} vs. \SII/\Ha \\ $\|$ \OIII/{\Hb} vs. \OII/\Ha} & 1,623 \\
\enddata
\end{deluxetable*}

The gas-phase extinction ($E(B-V)$) is calculated using the Balmer decrement, that is the change of the flux ratio of {\Ha/\Hb} or {\Hb/\Hg} relative to its intrinsic value. The intrinsic ratio ({\Ha/\Hb})$_0$ is 2.86 and (\Hb/\Hg)$_0$ is 2.137, under the Case B recombination for $T_e = 10^4$ K and electron density of $N_e = 100$ cm$^{-3}$ \citep{Hum87}. Here, we impose 3$\sigma$ cuts on {\Hb} and either {\Ha} or {\Hg} emission lines. We preferentially use the {\Ha/\Hb} to calculate the extinction if the {\Ha} line is not redshifted out of the wavelength coverage ($z\lesssim0.49$) and meets the SNR requirement. \boldtext{As a common practice, the $E(B-V)$ is set to zero when the observed line ratios are lower than the intrinsic values (about 32\% of the galaxies). It could be due to the uncertainties of the line ratio or statistical deviations from the theoretical values.} The median reddening is about 0.08 mag for the {\OIIIFOT}-selected galaxies. All the emission-line fluxes are corrected with the extinction using the attenuation law of \citet{Car89}. 

\begin{figure*}[ht!]
\centering
\includegraphics[width=1.0\textwidth]{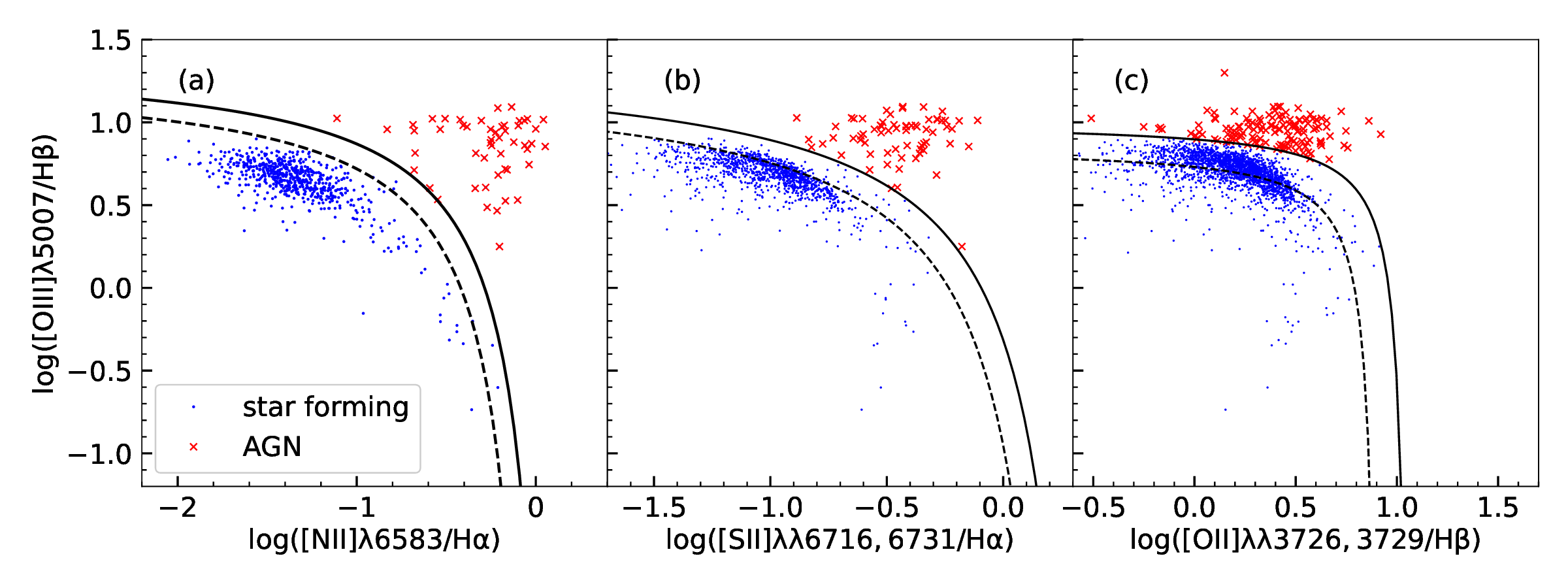}
\caption{BPT diagnostic diagrams to discriminate star-forming galaxies (blue dots) from AGNs (red crosses). These galaxies shown are \OIIIFOT-detected. (a) The BPT diagram of {\OIIIFIZ/\Hb} vs. {\NIIS/\Ha}. The dashed line shows the empirical division between starbursts and AGNs obtained by \citet{Kau03}. (b) The BPT diagram of {\OIII/\Hb} vs. {\SII/\Ha}. The dashed show the theoretical division between starbursts and AGNs obtained by \citet{Kew01}. The solid lines in these two panels display the demarcation curves with an additional uncertainty of 0.1 dex. (c) The galaxy distribution on the plot of {\OIIIFIZ/\Hb} vs. {\OII/\Hb}. The dashed line presents the separation limit proposed by by \citet{Lam10}. The solid line displays the separation limit with the addition of an extra 0.15 dex uncertainty, which is used as our new proposed demarcation between starbursts and AGNs.  \label{fig:bpt}}
\end{figure*}

Only metal-poor star-forming galaxies are the focus of this paper, so AGNs should be also excluded. Their optical spectra are dominated by non-thermal emission, making it complicated to study the stellar population. Some diagnostic diagrams with emission line ratios are usually used to discriminate different ionizing sources. Typically, the BPT diagrams \citep{Bal81} of {\OIII/\Hb} vs. {\NII/\Ha} and  {\OIII/\Hb} vs. {\SII/\Ha} are the standard tools to separate star-forming galaxies from AGNs. Figure \ref{fig:bpt} shows the method of using the BPT diagrams to select star-forming galaxies. The data points in this figure are all the \OIIIFOT-detected galaxies. In the diagram of {\OIIIFIZ/\Hb} vs. {\NIIS/\Ha} in Figure \ref{fig:bpt}a, the empirical starburst limit as shown in solid line is adopted from \citet{Kau03}:
\begin{equation}
{\rm \log\left(\frac{[OIII]\lambda5007}{H\beta}\right) = \frac{0.61}{\log([NII]\lambda6583/H\alpha)-0.15}+1.4}, 
\end{equation} 
where a modeling error of 0.1 dex \citep{Kau03} is added to the original formula. The diagnostic diagram of {\OIII/\Hb} vs. {\SII/\Ha} in Figure \ref{fig:bpt}b provides a supplementary classification, since the {\NIIS} line is not always detected and often blended with {\Ha}. The separation curve is proposed by \citet{Kew01}:
\begin{equation}
{\rm \log\left(\frac{[OIII]\lambda5007}{H\beta}\right) = \frac{0.72}{\log([SII]\lambda\lambda6716,6731/H\alpha)-0.42}+1.40},
\end{equation} 
where we also add an additional error of 0.1 dex \citep{Kew01}. The original separation line as shown in dashed curve in Figure \ref{fig:bpt}b is too close to the star-forming sequence. It would classify normal star-forming galaxies as AGNs. The above two BPT diagrams can only be used at redshifts $z<0.49$. At higher redshift, we attempt to use the diagnostic diagram of {\OIIIFIZ/\Hb} vs. {\OII/\Hb} first proposed by \citet{Lam04} and revised in \citet{Lam10}. Figure \ref{fig:bpt}c shows the galaxy distribution on this diagram. The dashed line in this plot shows the updated demarcation curve of \citet{Lam10} to separate starbursts and AGNs. The solid line shows the separation limit plus an uncertainty of 0.15 dex. \boldtext{From this figure, we can see that a significant fraction of galaxies in the star-forming regions of the former two BPT diagrams lying above the curve, so the \citet{Lam10} separation is not optimal especially for galaxies with small {\OII/\Hb} and large {\OIIIFIZ/\Hb}.} Hence, we adopt the demarcation curve with the addition of 0.15-dex uncertainty \citep{Lam10}, which is expressed as:
\begin{equation}
{\rm \log\left(\frac{[OIII]\lambda5007}{H\beta}\right) = \frac{0.11}{\log([OII]\lambda\lambda3726,3729/H\beta)-1.07}+1.00}. \label{equ:bpt3}
\end{equation} 
This new proposed separation limit is used to screen AGNs at higher redshifts in our sample via the the diagram of {\OIIIFIZ/\Hb} vs. {\OII/\Hb}. \boldtext{We should note that the additional errors are added to both ordinate and abscissa in Figure \ref{fig:bpt} in order to set looser selection cuts of star-forming galaxies and make discrimination curves more suitable to our data.} By using the above three BPT diagnostic diagrams, we get 1,623 star-forming galaxies. All the following analyses in this paper are based on this sample.


\section{Oxygen Abundance Determination and Other Property Measurements} \label{sec:measure}
\subsection{Direct measurement of oxygen abundance}
The gas-phase metallicity can be estimated using the calibrations developed from either theoretical models or empirical methods. The direct method to measure the metallicity is the so-called {\Te} method that relies on the determination of the electron temperature (\Te) from auroral emission lines. {\Te} can be measured by using the line intensity ratio of two different ionization species of the same element,  such as \OIII$\lambda\lambda$4959,5007/\OIII$\lambda$4363 \citep{All84}. Although the direct $T_e$ method is subject to some caveats, such as metallicity underestimation caused by possible temperature fluctuation or gradients in high-metallicity star forming regions \citep{Kew08,md23}, it is currently the most reliable and accurate method to measure the metallicity.

In the direct $T_e$ method, the two-zone model is assumed, where the low and high ionization zones are traced by low and high ionization emission lines, respectively. The electron temperature in the high ionization zone $T_e$(\OIII) can be estimated using the line ratio of {\OIIIFOT} and \OIII$\lambda\lambda$4959,5997, given the electron density (\Ne) that can be calculated with the {\OII} or {\SII} doublets. The electron temperature in the low ionization zone $T_e(\OII)$ can be traced by \OII$\lambda$7320,7330 lines or \NII$\lambda$5755, which are too weak to be detected in the DESI spectra. An iterative method from \citet{Nic14} is adopted to determine $T_e$(\OII) and the total oxygen abundance:

\begin{equation}
\begin{split}
&\Te (\OII) = \Te(\OIII) \times \\
& (3.0794 - 0.086924Z - 0.1053Z^2 + 0.010225Z^3), \label{equ:OII}
\end{split}
\end{equation}
where $Z$ is the oxygen abundance. The iteration process starts with the O$^{++}$ abundance and the total oxygen abundance is the summation of the O$^+$ and O$^{++}$ abundances determined by the two-zone electron temperatures. Usually, the calculation converges within several iterations.  

The {\SII} doublet (6716 and 6731 \AA) can be detected for galaxies at $z<0.45$, while at higher redshift we can use the \OII$\lambda\lambda$3726,3729 lines to calculate the electron density. We require that the flux SNRs of these lines are greater than 4. If the SNRs are not high enough or the line ratios are out of the valid ranges to determine \Ne, we assume a constant {\Ne} $= 100 $ cm$^{-3}$ \citep{And13,LyC14,Gao17}. The electron temperature is insensitive to {\Ne} in the relatively low density regime and assuming {\Ne} $ = 100$ cm$^{-3}$ is consistent with the measurements from the {\OII} or {\SII} doublet. We use the Python package PYNEB\footnote{\url{http://www.iac.es/protecto/PyNeb/}} developed by \citet{Lur15} to calculate the electron density, temperature, and oxygen abundance. As mentioned in \citet{Gao17}, the atomic recombination and collision strength data adopted in PYNEB are outdated, with the consequence that the electron temperature would be overestimated and thus the metallicity underestimated. We replace them with the updated data for O$^+$, O$^{++}$, and S$^+$ from \citet{Fro04,Kis09,Tay10,Sto14}. If the {\SII} or {\OII} doublets have enough SNR, the module \textit{getCrossTemDen} in PYNEB is used to calculate {\Ne} and \Te(\OIII) simultaneously. If {\Ne} is set to 100 cm$^{-3}$, the PYNEB module \textit{getTemDen} is used to calculate \Te(\OIII). \boldtext{The abundance of O$^+$/H$^+$ is determined by the PYNEB module \textit{getIonAbundance} with \OIII$\lambda\lambda$4959,5007/{\Hb} and \Te(\OIII). The O$^{++}$/H$^+$ abundance is obtained by the same PYNEB module with \OII$\lambda\lambda$3726,3929/{\Hb} and \Te(\OII) estimated by Equation (\ref{equ:OII}).} The total oxygen abundance 12+log(O/H) is calculated by adding the two abundances. The error estimation is implemented through Monte Carlo simulations. \boldtext{We add random errors to the fluxes of the measured emission lines according to their esitmated uncertainties. This process is repeated 1000 times.} Each combination of the fluctuated fluxes is fed to PYNEB to calculate \Ne, \Te, and 12+log(O/H).  The final value of each parameter is calculated as the median value and the error is estimated from the 68\% confidence interval of the parameter distribution. 

\begin{figure*}[tbh!]
\includegraphics[width=1.0\textwidth]{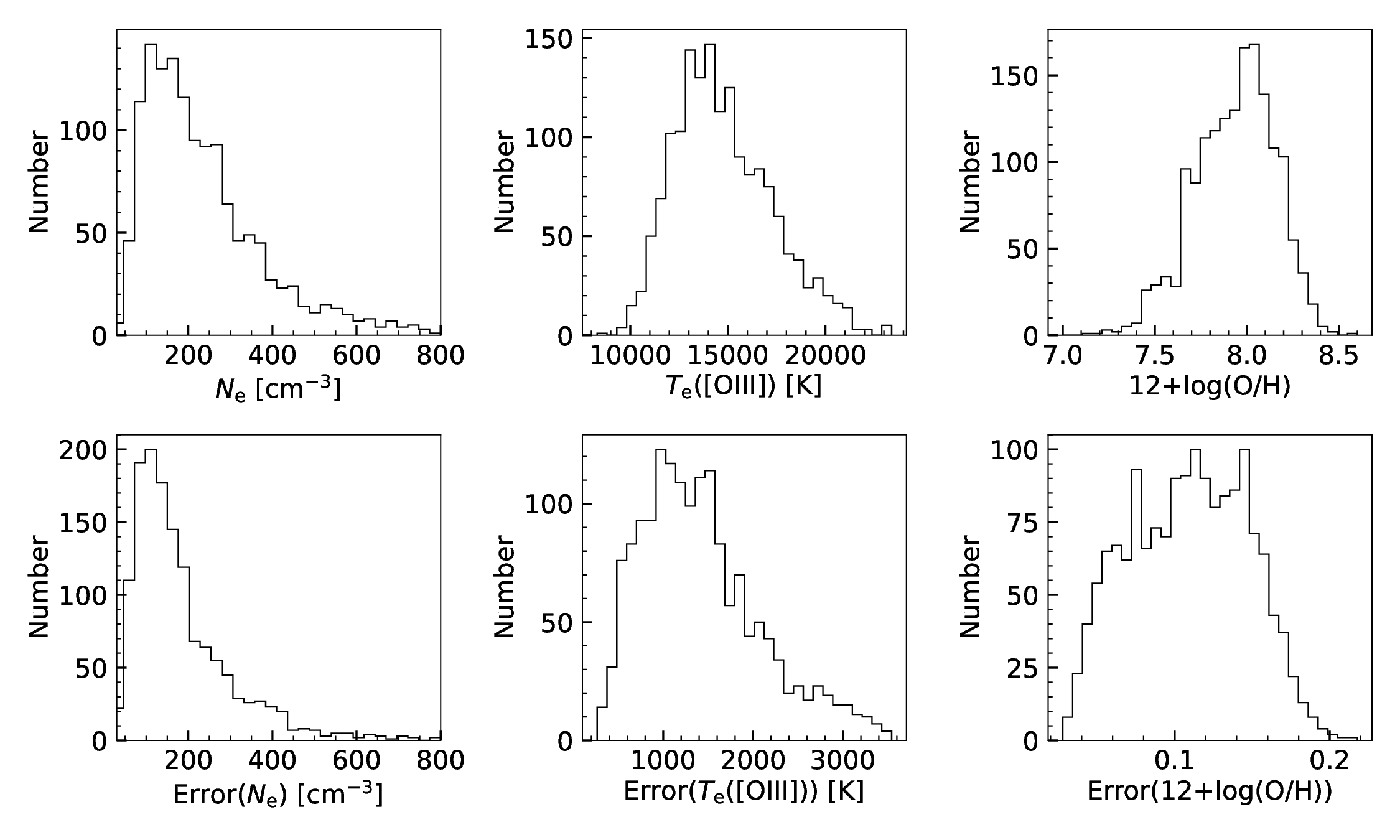}
\caption{\boldtext{Top panels: distributions of {\Ne} (left), \Te(\OIII) (middle), and 12+log(O/H) (right) of our \OIIIFOT-selected galaxies. Bottom panels: error distributions of {\Ne} (left), \Te(\OIII) (middle), and 12+log(O/H) (right).} \label{fig:dentemp}}
\end{figure*}

Figure \ref{fig:dentemp} shows the distributions of {\Ne}, \Te(\OIII) and 12+log(O/H) \boldtext{and their error distributions}. There are a total of 702 and 692 galaxies with {\Ne} calculated by {\SII} and {\OII}, respectively. The remaining 229 galaxies have the assumed value of $\Ne=100$ cm$^{-3}$. About 90\% of the galaxies have an electron density in the range of 76--629 cm$^{-3}$. The assumed $\Ne=100$ cm$^{-3}$ in this paper is in the valid range. There are about 1,608 galaxies with effective {\Te} measurement. The electron temperature of 90\% of the galaxies ranges from 11,272 to 19,559 K and the median value is 14,421 K. The median error for {\Ne} and  \Te(\OIII) is 150 cm$^{-3}$ and 1,326 K. Although the relative error of {\Ne} is large, it has little effect on the determination of the oxygen abundance due to its insensitivity to the electron density. The oxygen abundance of 90\% of galaxies in our \OIIIFOT-selected galaxies ranges from 7.55 to 8.26 dex and the median value is 7.96 dex. The median metallicity error is 0.11 dex.

\subsection{Stellar mass estimation}
The stellar mass of each galaxy is estimated through stellar population synthesis fitting using the multi-wavelength photometric data, as well as the DESI optical spectrum. The photometric data come from the data release DR9 of the DESI legacy imaging surveys (hereafter, LS-DR9), which are mainly used for the target selections of the DESI spectroscopic survey \citep{Zou17, Dey19}. The imaging data in LS-DR9 provide both optical and infrared photometry, including 3 three optical bands ($g,r$, and $z$) and 4 infrared bands ($W1$, $W2$, $W3$, and $W4$) from WISE. Due to much shallower photometric depths and lower spatial resolutions of $W3$ and $W4$ bands, we only use the $W1$ and $W2$ to construct the photometric SED. The limiting magnitudes in $g$, $r$, $z$, $W1$, and $W2$ are 24.7, 23.9, 23.0, 20.7, and 20.0 mag, respectively \citep{Zou22}. Although the WISE imaging is relatively shallow and has a low resolution of about 6{\arcsec} (by contrast, 0.9{\arcsec} for $z$ band),  LS-DR9 provides consistent force-modeling photometry for both optical and infrared bands. Considering the importance of the near-infrared bands in estimating the stellar mass, we require that the two WISE bands have effective photometry with SNR $>2$.  

The DESI optical spectra, on the other hand, cover the wavelength range of 3650--9800 \AA, providing complementary information about the stellar population. However, the continuum SNR especially for ELGs is low, so that we can hardly constrain the stellar mass by only using full spectral fitting techniques. Instead, we obtain artificial broad-band spectrophotometry by convolving the optical spectra with self-defined 10 contiguous broad bands, which start from 3,650 {\AA} and have bandwidths of about 615{\AA}. Each self-defined filter is a boxcar filter with a flat transmission in a window of 615 {\AA} width. Before convolution, we first scale each galaxy spectrum by multiplying by a scaling factor. The scaling factor is calculated as an aperture correction by comparing the total photometric fluxes in the $r$ band with the spectrophotometric flux in this filter. \boldtext{Here we assume that the light collected by the fiber is proportional to the total light of the galaxy and the color of each galaxy is the same inside and outside of the fiber.} The 10 customized broad-band spectrophotometry as well as the 5 broad-band photometry, are used for determining the stellar mass. \boldtext{Both 5-band photometry and spectroscopic based photometry are used together to infer stellar masses.} We should also note that in some cases the DESI targets are not selected from the DESI imaging surveys, in which case the spectra are scaled to the the Gaia $G$-band photometry. 

\begin{figure*}[ht!]
\includegraphics[width=\textwidth]{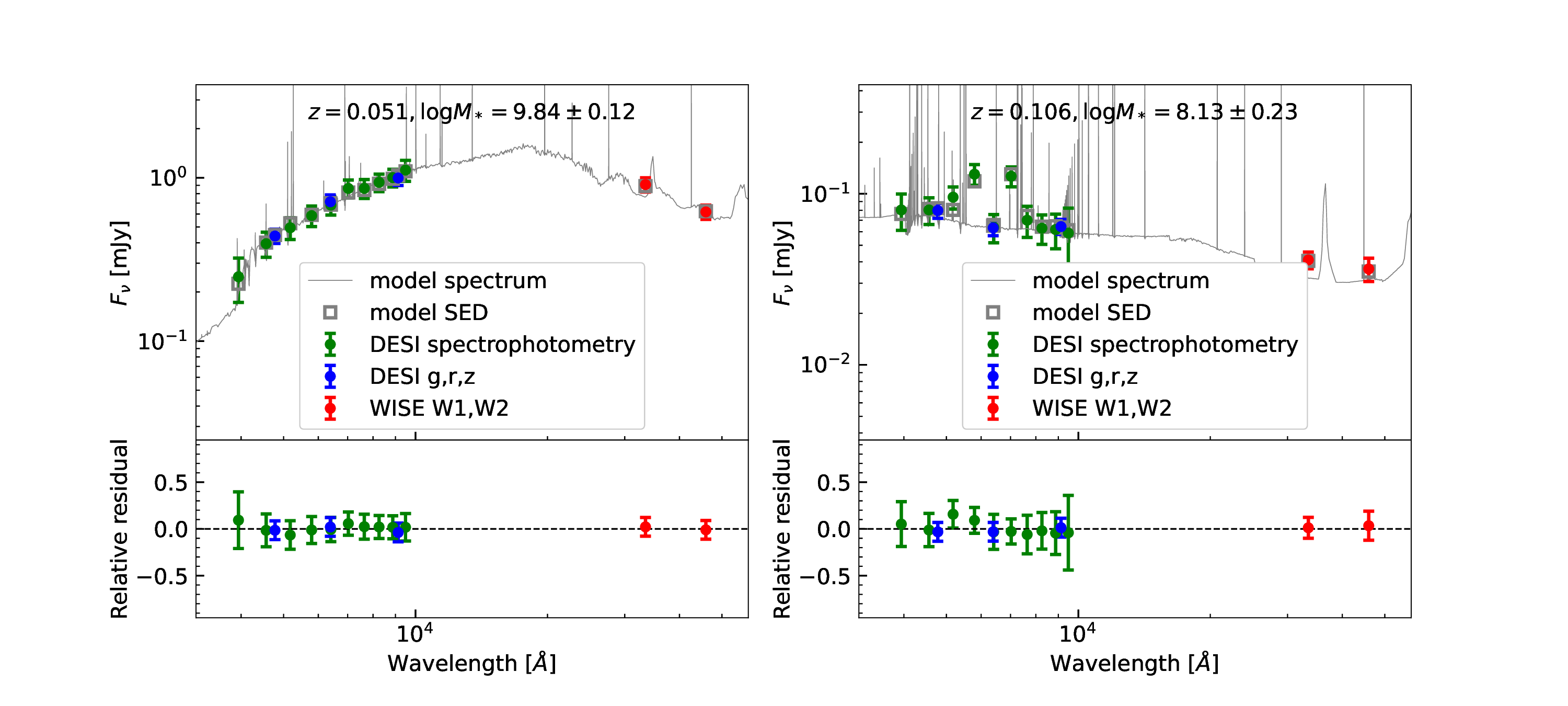}
\caption{Two examples of the CIGALE stellar population synthesis fitting based on the DESI photometric and spectrophotometric data (left for an old quiescent galaxy at $z\sim0.05$ and right for a young star-forming galaxies at $z\sim0.11$). The redshift and derived stellar mass for each galaxy are annotated in the top panel. The solid points with error bars represent the observed SEDs. The grey lines are the best-fit model spectra and the grey open squares display the model SEDs. The bottom panel shows the relative residual between observed and model SEDs, which is $(F_\mathrm{obs}-F_\mathrm{model})/F_\mathrm{obs}$, where $F_\mathrm{obs}$ and $F_\mathrm{model}$ are the observed and model fluxes for a given filter, respectively. The blue points with error bars show the DESI optical photometric fluxes and corresponding errors in $g, r$, and $z$ bands, while the red ones display the WISE W1 and W2-band photometry. The green points are 10  spectrophotometric fluxes derived by \boldtext{convolving the DESI spectrum with 10 virtual broad-band filters and scaled up to correct for aperture effects.} \label{fig:sedfit}}
\end{figure*}

We use the version 2022.1 of the Code Investigating GALaxy Emission \citep[CIGALE;][]{Boq19,Yan20,Yan22} to model the spectral energy distributions (SEDs) of galaxies. The CIGALE code can efficiently reconstruct the galaxy SED ranging from the far-ultraviolet to the infrared with highly flexible modeling of composite stellar population, gas emission, and dust attenuation and radiation. This modeling returns important physical properties of the galaxies, including stellar mass, star formation rate (SFR), dust extinction, and stellar age. In CIGALE, we adopt the  simple stellar population (SSP) models of \citet{Bru03}, a \citet{Cha03} initial mass function (IMF), and a delayed star formation history (SFH). We select 21 stellar population ages (1 Myr -- 13 Gyr), five stellar metallicities (0.0004, 0.004, 0.008, 0.02, 0.05), and 14 star formation timescales $\tau$ for the SFH ranging from 1 Myr to 80 Gyr.  The \citet{Cha00} extinction law and \citet{Dra14} dust IR emission models are adopted to account for the dust attenuation of the interstellar medium and possible emission excess in WISE W1 and W2 bands, respectively. Due to strong emission lines in the spectra caused by the ionization of high-mass star radiation, the nebular emission is taken into account. However, no AGN emission is considered in the SED modeling, as we have already excluded those galaxies with AGN. The redshift is fixed to the spectroscopic redshift when constructing the models. Figure \ref{fig:sedfit} shows two examples of the SED fitting with CIGALE to present two typical galaxies with old and young stellar populations, respectively.

\begin{figure}[ht!]
\centering
\includegraphics[width=1.0\columnwidth]{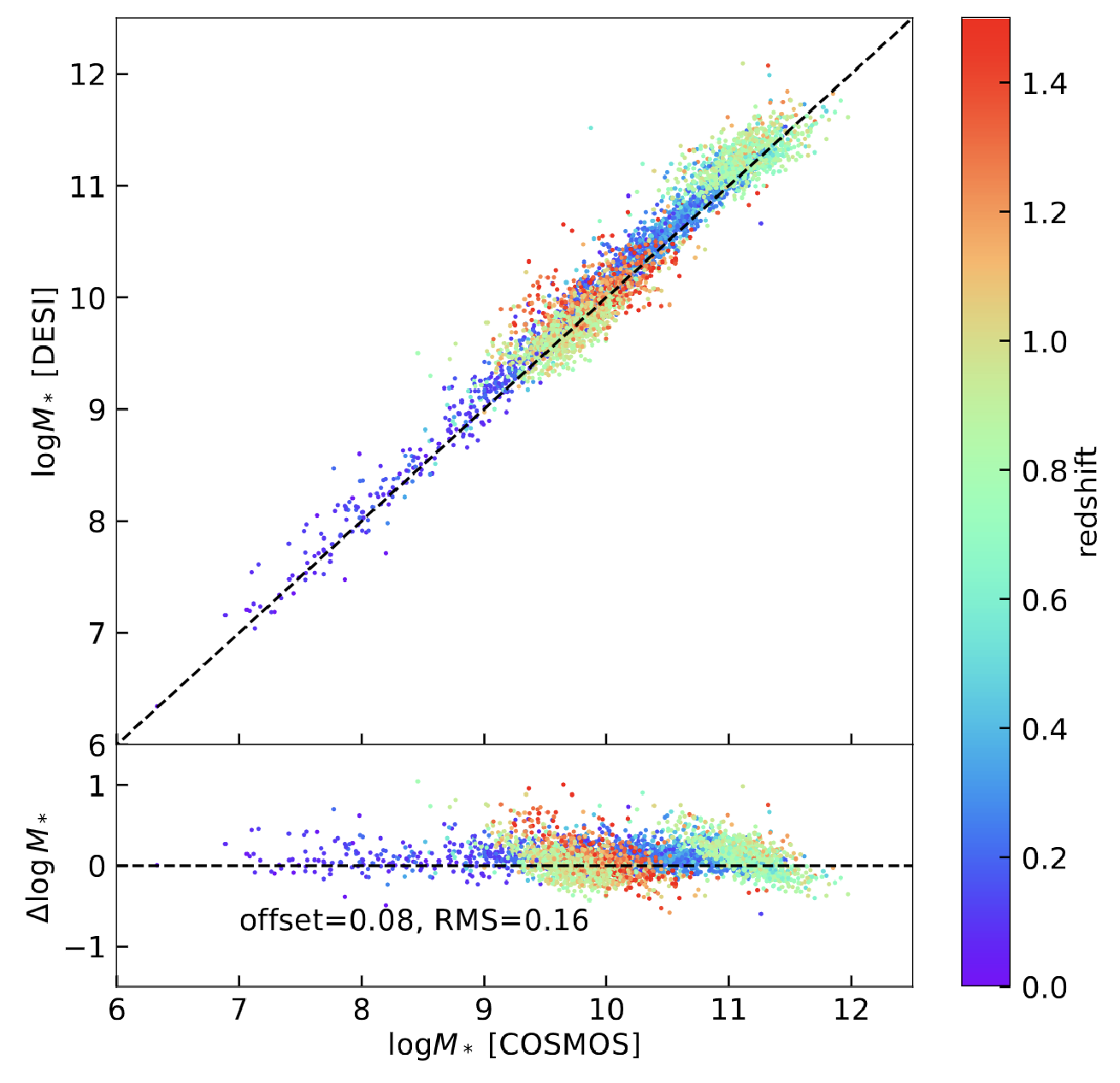}
\caption{Comparison of the stellar masses from DESI and COSMOS2020.  The top panel presents the stellar mass comparison and the diagonal line represents the exact match of the two measurements. The bottom panel shows the mass difference of these two surveys as function of the stellar mass in the COSMOS2020 catalog. The horizontal line represents $\Delta\log{M_*}=0$. The data points are color-coded by the DESI spectroscopic redshift. The median \boldtext{difference} and scatter of $\Delta\log{M_*}$ are also displayed in the bottom panel. \label{fig:masscomp}}
\end{figure}

To assess the quality of our stellar mass measurements, we cross-match the entire DESI galaxy sample with the COSMOS2020 catalog compiled by \citet{Wea22} for the Cosmic Evolution Survey (COSMOS). They have performed multi-wavelength photometry (35 bands) and calculated photometric redshifts for galaxies down to the depth of $i\sim27$ over a 2 deg$^2$ footprint. The \textit{Farmer} catalog and stellar mass derived with the LePhare SED fitting tool is used in this paper. There are more than 6300 galaxies observed by DESI located in the COSMOS field. We limit the galaxies for comparison with the photometric redshift accuracy ($|\Delta z|=|(z_\mathrm{phot}-z_\mathrm{spec})/(1+z_\mathrm{spec})|$, where $z_\mathrm{phot}$ and $z_\mathrm{spec}$ are photometric and spectroscopic redshifts, respectively) less than 0.02 and the logarithmic stellar mass uncertainty less 0.1 dex \boldtext{(artificially set to relatively smaller values for better comparison)}, generating a set of about 3600 galaxies. Figure \ref{fig:masscomp} shows the comparison of the stellar mass calculated using the DESI and COSMOS data. It can be seen that the stellar mass measurements of these two surveys are consistent.  There is a median bias of about 0.08 dex for $\Delta\log{M_*} = \log{M_*}[\mathrm{DESI}]-\log{M_*}[\mathrm{COSMOS}]$ . It is mainly caused by different configurations in the SED fitting codes for constructing galaxy spectral models \boldtext{(e.g. different IMFs, model spectral libraries, star formation histories).} The dispersion $\sigma_{\Delta\log{M_*}}$ is about 0.16 dex, which is comparable to the estimated uncertainty with the DESI data (median of 0.18 dex). We also compare the stellar mass derived with only 5-band photometry and find that it has a similar bias but slightly larger dispersion of $\sigma_{\Delta\log{M_*}}\sim0.17$ dex. Thus, the photometry plus spectrophotometry can indeed improve the stellar mass estimation. In particular, the spectrophotmetry is indispensable for those galaxies without DESI LS photometry.

\subsection{Star formation rate}
The star formation rate is calculated from the de-reddened {\Ha} or {\Hb} luminosity using the Kennicutt relation \citep{Ken98}: 
\begin{equation}
\mathrm{SFR} = 7.9\times10^{-42} L(\Ha) = 7.9\times10^{-42} F(\Ha) 4\pi d_L^2, \label{equ:sfr}
\end{equation}
where $F(\Ha)$ is the extinction-corrected {\Ha} flux, $d_L$ is the luminosity distance, and $L(\Ha)$ is the {\Ha} luminosity. We preferentially use the {\Ha} luminosity to estimate the SFR. At $z>0.49$ where the {\Ha} line is redshifted out of the wavelength coverage, the SFR from the dereddened {\Hb} luminosity is calculated assuming the intrinsic flux ratio of $(\Ha/\Hb)_0= 2.86$. Note that the emission-line luminosity is aperture-corrected using the same scaling factor as the one when we estimate the stellar mass through SED fitting. The SFR error is mainly from the calibration uncertainty of Equation (\ref{equ:sfr}) (about 30\%). Note that the uncertainty from the aperture correction is not included.


\section{Our identified XMGPs and Their Properties}\label{sec:xmpgs}
\subsection{XMPGs identified from $T_e$-based Oxygen Abundance}
From the oxygen abundance distribution in Figure \ref{fig:dentemp}, we can see that 902 \OIIIFOT-selected galaxies are metal-poor galaxies, which are defined to have 12+log(O/H) $<7.99$ ($Z < 0.2\Zsun$). No galaxy has an oxygen abundance larger than the solar metallicity. There are 223 XMPGs with 12+log(O/H) $<7.69$ ($Z < 0.1\Zsun$). These XMPGs account for 13.9\% of all the \OIIIFOT-detected galaxies. They are about 0.01\% of the total DESI observed galaxies, indicating their rarity. \boldtext{We define those galaxies with $Z+E(Z) <0.1\Zsun$ as \textit{confirmed XMPGs} and those with $Z < 0.1\Zsun$ but $Z+E(Z) \ge 0.1\Zsun$ as \textit{XMPG candidates}, where $E(Z)$ is the abundance uncertainty.} There are 95 confirmed XMPGs, 13 and 2 of which have metallicity less than 1/20 and 1/30 \Zsun, respectively. The remaining 134 galaxies are XMPG candidates. \boldtext{Only six of our XMPGs are previously identified, if a cross-matching radius of 2 arcsecs is considered.}  

\begin{figure*}[tbh!]
\centering
\includegraphics[width=1.0\textwidth]{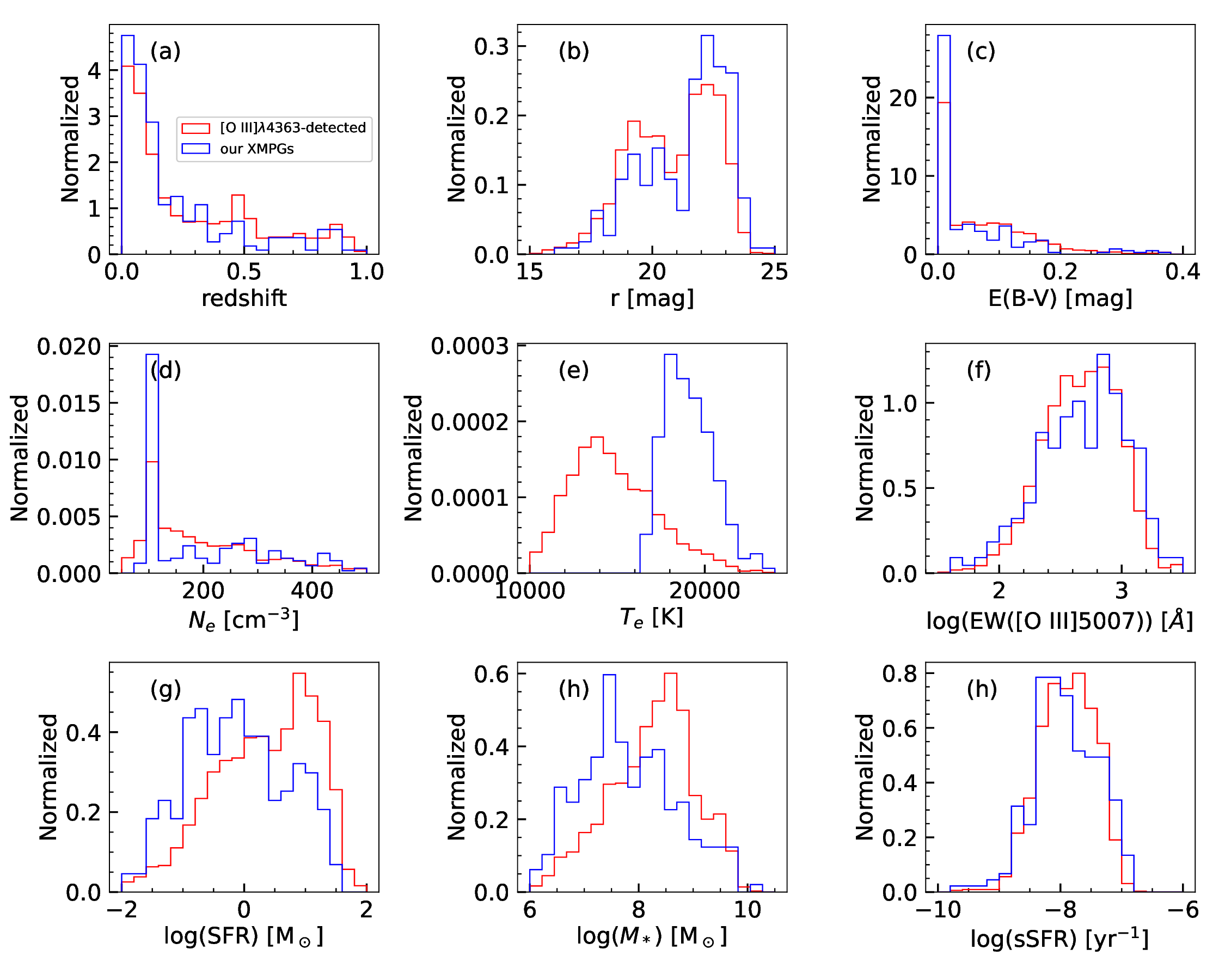}
\caption{Distributions of basic properties for all our XMPGs (blue) and \OIIIFOT-detected galaxies (red), including redshift (a), $r$-band magnitude in mag (b), gas-phase extinction $E(B-V)$ in mag (c), electron density $N_e$ (d) in cm$^{-3}$, electron temperature $T_e$ in K (e), logarithmic equivalent width of \OIII$\lambda$5007 in {\AA} (f),  logarithmic SFR in \Msun yr$^{-1}$ (g), logarithmic stellar mass in {\Msun} (h), and logarithmic sSFR in yr$^{-1}$ (i). \label{fig:mpgprops}}
\end{figure*}

Figure \ref{fig:mpgprops} shows the distributions of basic properties of all our XMPGs. About 74\% of the XMPGs are located at relatively low redshift of $z < 0.3$.  There are 58 $z>0.3$ XMPGs with the highest redshift up to 0.96.  This is by far the largest sample of high-redshift XMPGs. The magnitude histogram in Figure \ref{fig:mpgprops}b presents a bimodal distribution, which corresponds to the two types of DESI spectroscopic targets (brighter for BGS and fainter for ELG). Our XMPGs are almost dust-free, and the median $E(B-V)$ is about 0.01 mag. The median $N_e$ and $T_e$ are 182 cm$^{-3}$ and 18,927 K, respectively. The median {\OIIIFIZ} equivalent width (EW(\OIIIFIZ)) is about about 528 \AA. The median SFR, stellar mass ($M_*$), and specific star formation rate (sSFR $=$ SFR/$M_*$) are 0.65 {\Msun} yr$^{-1}$, 4.65$\times10^7 \Msun$, and $1.14\times10^{-8}$ yr$^{-1}$. About 89\% of our XMPGs are dwarf galaxies with $M_* < 10^9 \Msun$. 

For comparison, we also present the parameter distributions of all \OIIIFOT-detected sample in Figure \ref{fig:mpgprops}. We notice that the XMPGs and the parent sample have similar distributions of redshift, magnitude, extinction, electron density, and EW(\OIIIFIZ). The XMPGs have lower metallicities and hence high $T_e$. From the SFR and mass distributions, we find that the XMPGs tend to present relatively lower SFRs than the parent sample, mostly due to their lower mass. Further investigations show that the sSFR distributions of these two samples are more similar. The median sSFR of XMPGs is around 1.14 $\times 10^{-8}$ yr$^{-1}$, which is comparable to that of other \OIIIFOT-detected galaxies (1.34 $\times 10^{-8}$ yr$^{-1}$), but about 8 times higher than that of normal star-forming galaxies (the control sample selected in Section \ref{sec:env}), whose median sSFR $\sim 1.36 \times 10^{-9}$ yr$^{-1}$. Metal-poor galaxies are undergoing intense star formation. Table \ref{tab:xmpgs} and \ref{tab:xmpgs_can} in the appendix list the confirmed XMPGs and XMPG candidates identified in this work, respectively.

\subsection{The most metal-poor galaxies} \label{sec:mostoh}
\begin{figure}[tbh!]
\centering
\includegraphics[width=1.0\columnwidth]{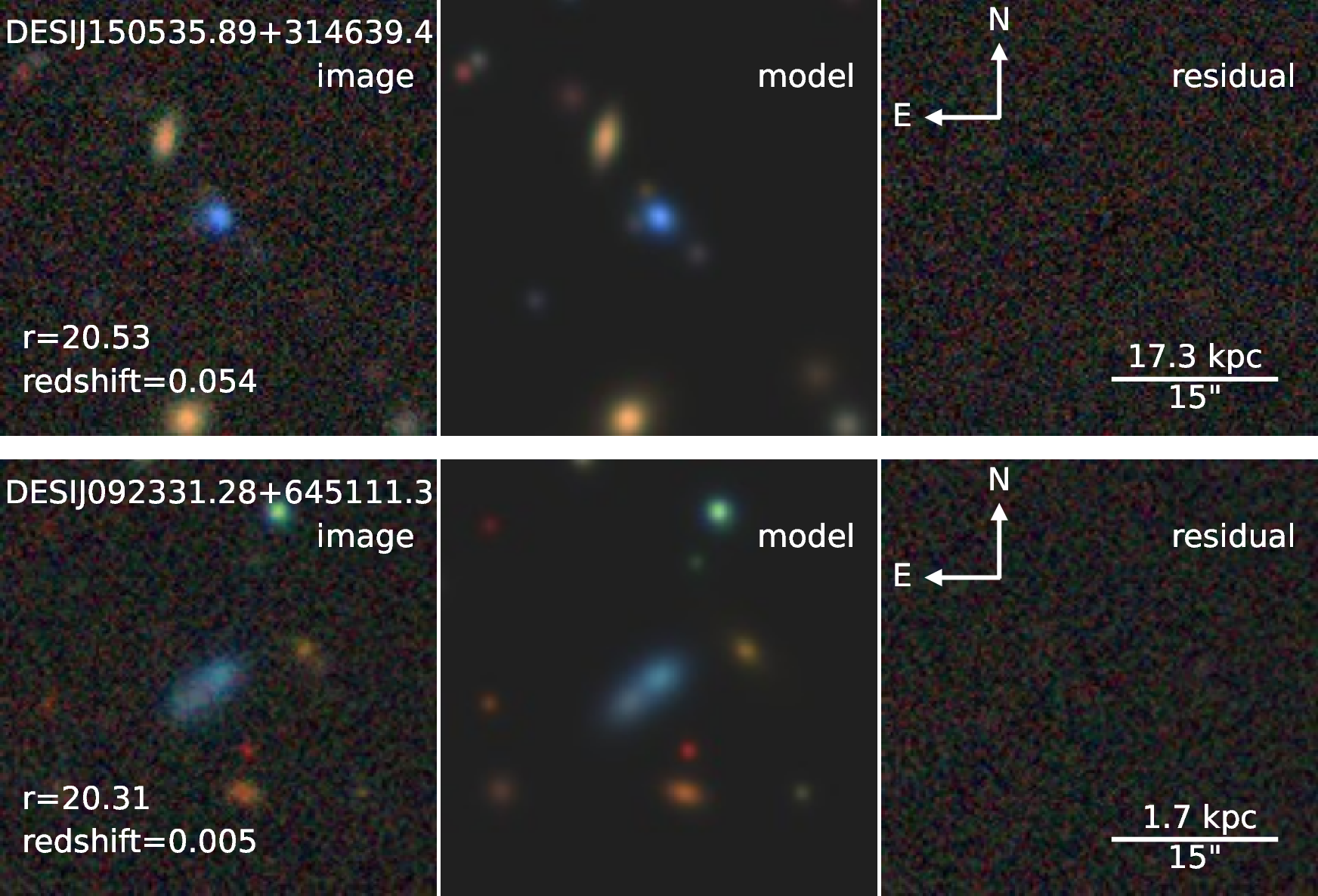}
\caption{Two most metal-poor galaxies identified in this work. The left, middle and right panels display the observed images, models, and residuals of the detected objects in the DESI legacy imaging surveys, respectively. The residual is the subtraction of the model from the observed image, where the model is derived by \boldtext{the Tractor photometry code via forward modeling of the galaxy profile \citep{Dey19}. The Tractor modelling provides the total integrated fluxes for galaxies.} The top panels show the images of DESIJ150535.89+314639.4 and the bottom ones present those of DESIJ092331.28+645111.3. The $r$-band magnitude and redshift are marked in the left panels. The horizontal line denotes the scales in arcsec and kpc at the distance of each galaxy. North is up and east is left.  \label{fig:mostxmpg}}
\end{figure}

\begin{figure}[tbh!]
\centering
\includegraphics[width=1.0\columnwidth]{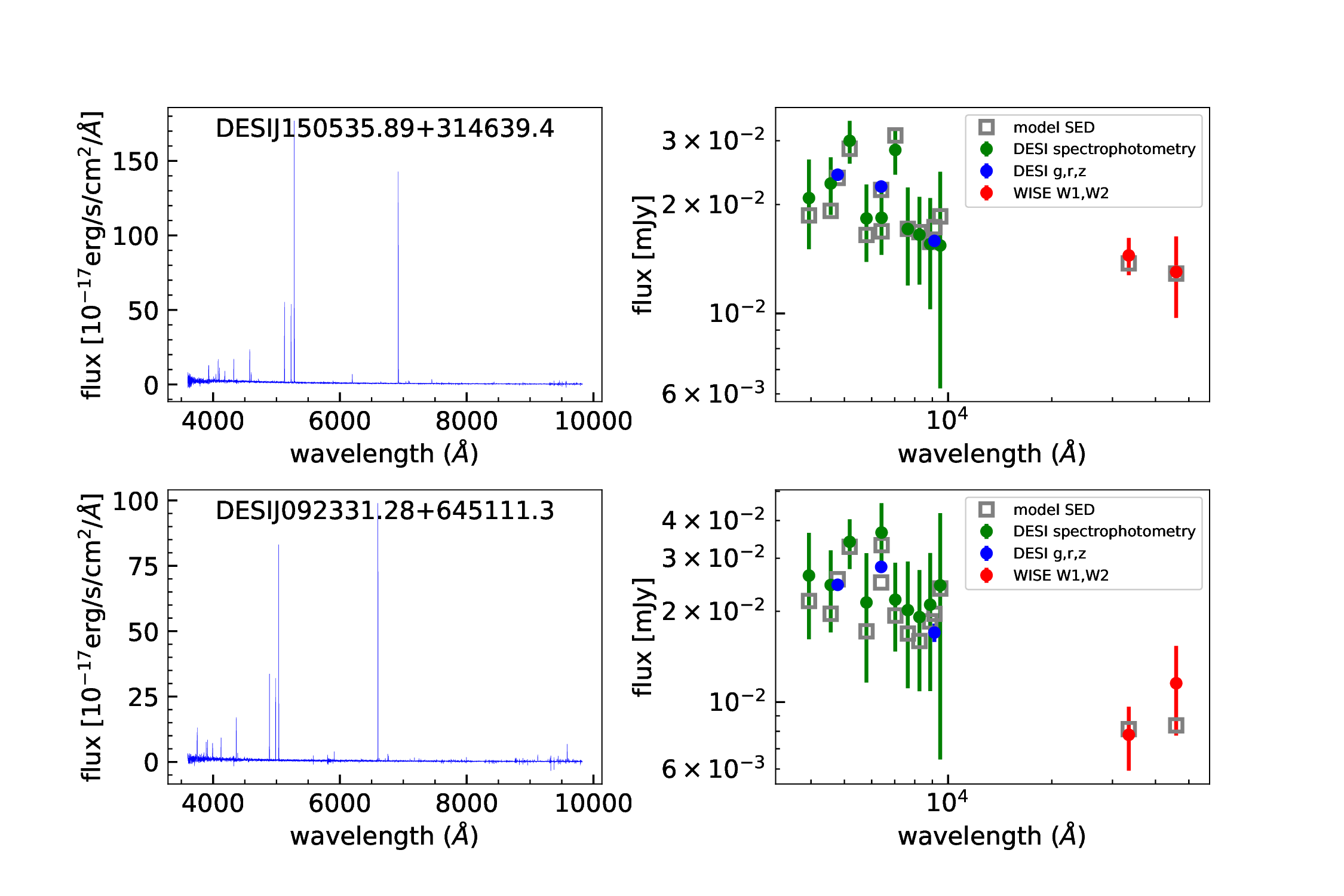}
\caption{Observed spectra (left panels) and multi-wavelength SEDs (right panels) for our two most metal-poor galaxies. The top panels are for DESIJ150535.89+314639.4 and the bottom ones are for DESIJ092331.28+645111.3. The colored symbols are the same as those in Figure \ref{fig:sedfit}. \label{fig:specsed}}
\end{figure}

The two most metal-poor galaxies in our sample, with an oxygen abundance $Z/\Zsun<1/30$, are: DESIJ150535.89+314639.4 at $z=0.054$ and DESIJ092331.28+645111.3 at $z=0.005$. Figure \ref{fig:mostxmpg} shows their color images composed with the $g$, $r$, and $z$-band imaging. The DESI photometric reduction pipeline implements a forward modeling for each object with a point spread function (PSF) profile or a galaxy profile \citep{Dey19}. The intrinsic shape parameters used in this paper are from the Legacy Survey DR9 \citep{Sch23}. The model and residual images in Figure \ref{fig:mostxmpg} show that the modeling works very well. The significant morphological discrepancy of these two XMPGs as shown in Figure \ref{fig:mostxmpg} indicates possible different origins to their low metallicities. Figure \ref{fig:specsed} shows the observed spectra and photometric SEDs of these two XMPGs. Strong {\OIIIFIZ} and relatively weak {\OII} emission lines in the spectra indicate that they have the high-ionization environments. Strong {\Ha} emission lines suggest intense star formation.

\textbf{DESIJ150535.89+314639.4} This is the most metal-poor galaxy (DESI TARGETID $=$ 39628517750604871) identified by us. The oxygen abundance is 12+log(O/H) $=7.16\pm0.05$ ($Z \sim 2.9\%\Zsun$). The lowest metallicity reported so far is about 1.6\%{\Zsun} \citep{Koj20}. DESIJ150535.89+314639.4 is located at the redshift of $z=0.054$. The $g$-band magnitude is 20.45 mag and the absolute magnitude is $M_g=-16.43$ mag. The optical colors are $(g-r)=-0.081$ and $(r-z)=-0.37$. The stellar mass and SFR are $1.51\times10^7\Msun$ and 0.22 {\Msun} yr$^{-1}$, respectively, giving a sSFR of $1.43\times10^{-8}$ yr$^{-1}$. From Figure \ref{fig:mostxmpg}, we can see that DESIJ150535.89+314639.4 is a blue compact dwarf (BCD) galaxy with the half-light radius $r_\mathrm{eff} = 640$ pc. The mean surface brightness $\mu_g$ within the half-light radius is about 21.22 mag arcsec$^{-2}$. It falls in the color cuts defined by \citet{Hsy18} to select BCD candidates and conforms to the BCD observational criteria listed in \citet{Mor11}. 

\textbf{DESIJ092331.28+645111.3} This is the second most metal-poor galaxy (DESI TARGETID $=$ 39633440282250119). The metallicity is 12+log(O/H) $=7.20\pm0.09$ ($Z\sim 3.2\%\Zsun$). It is located at $z=0.0054$ and $(\alpha, \delta)=(140.88032\arcdeg, 64.85315\arcdeg)$. We can see from Figure \ref{fig:mostxmpg} that DESIJ092331.28+645111.3 has a complex morphology, exhibiting a tadpole-like or cometary shape. Such a head-tail morphology is common among XMPGs as shown in \citet{Mor11, San16a, Koj20}. The DESI fiber pointed to the head (i.e. the north-west part). The $g$-band apparent and absolute magnitudes of the head are $g=20.46$ and $M_g=-11.38$. The optical colors are $(g-r)=0.15$ and $(r-z)=-0.53$.  The stellar mass and SFR are $1.38\times10^5\Msun$ and 0.0034 {\Msun} yr$^{-1}$, respectively, giving sSFR $= 2.42\times10^{-8}$ yr$^{-1}$. If we take the south-east part into account and assume the same mass-luminosity ratio (M/L) as the north-west one, the total stellar mass would be 1.67 times larger (may be larger due to possibly higher M/L for the tail). The sSFR is slightly larger than DESIJ150535.89+314639.4, suggesting that its mass assembly is faster although lower mass. Nevertheless, this extremely metal-poor head shows a blue color and compact size ($r_\mathrm{eff}=172$ pc). It connects to an extended fainter tail with a lower surface brightness. It was found that the XMPG heads and tails are dynamically connected and the tails have older stellar populations and are more metal-rich than the heads \citep{San15}.

DESIJ150535.89+314639.4 and DESIJ150535.89+314639.4 present two different morphologies, suggesting that they might have distinct physical origins. It is also possible that DESIJ150535.89+314639.4 might be in the stage of bursty star formation, so that the low surface brightness tail is outshone by the bright starburst clump. \boldtext{Very faint structures can be seen around this galaxy. To check these, we need higher-resolution and deeper images.} We will systematically investigate the morphologies and stellar population properties of our XMPG sample in the future.

\subsection{XMPG distributions on the color-color and BPT diagrams}
\begin{figure}[tbh!]
\includegraphics[width=1.0\columnwidth]{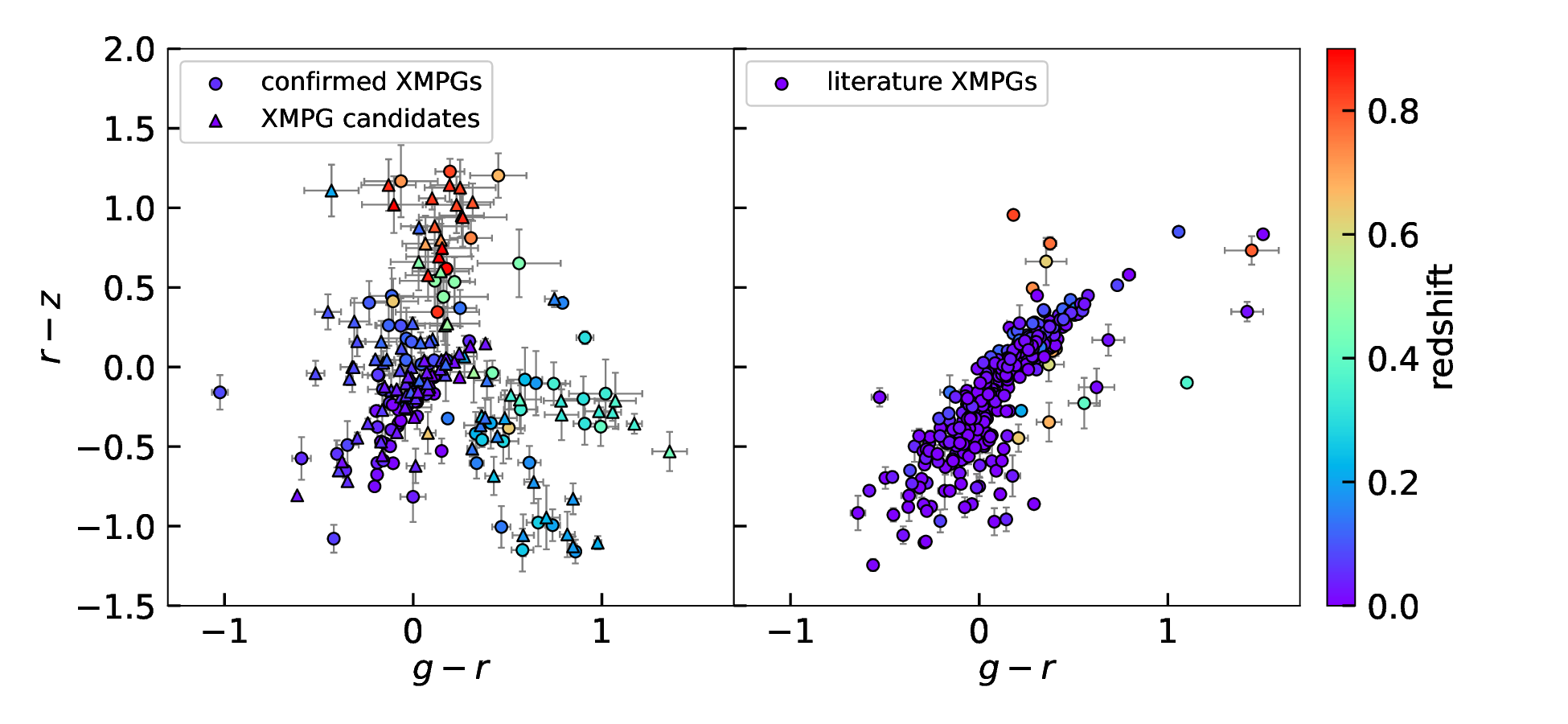}
\caption{\boldtext{Observed color-color diagrams} for our DESI XMPGs (left) and the XMPGs collected from the literature (right). The points in each panel are color-coded by redshift. The grey error bars show the uncertainties of the colors. In the left panel, the circles represent our confirmed XMPGs, while the triangles represent the XMPG candidates. \label{fig:ccd}}
\end{figure}

The number of XMPGs identified here with DESI is almost double what is currently known. There are 338 XMPGs collected from the literature that are cross-matched with the DESI photometric catalogs (the photometric errors in $g$, $r$, and $z$ bands are limited to be less than 0.2 mag).  Figure \ref{fig:ccd} compares the distributions of the DESI (this work) and literature XMPGs on the color-color diagram of $g-r$ vs. $r-z$. \boldtext{The literature XMPGs are collected by Sui et al. (in preparation).} The magnitudes have been corrected for the Galactic extinction. In the right panel of this figure, we can see that most XMPGs in the literature seem to lie in a low-redshift sequence. Only a few XMPGs are identified at redshift of $z>0.3$. Our local XMPGs at $z<0.1$ also follow this sequence. A large sample of relatively higher-redshift XMPGs are identified to the right and the top of the sequence. This is mainly caused by the strong {\Ha} and {\OIIIFIZ} lines redshifting into or out of the $r$ and $z$ band, \boldtext{which is similar to the effect of green pea galaxies \citep{Car09}.} These galaxies and their positions in the color space will help to identify XMPG candidates at $z<1$ from future large-scale imaging surveys.

\begin{figure}[htb]
\centering
\includegraphics[width=1.0\columnwidth]{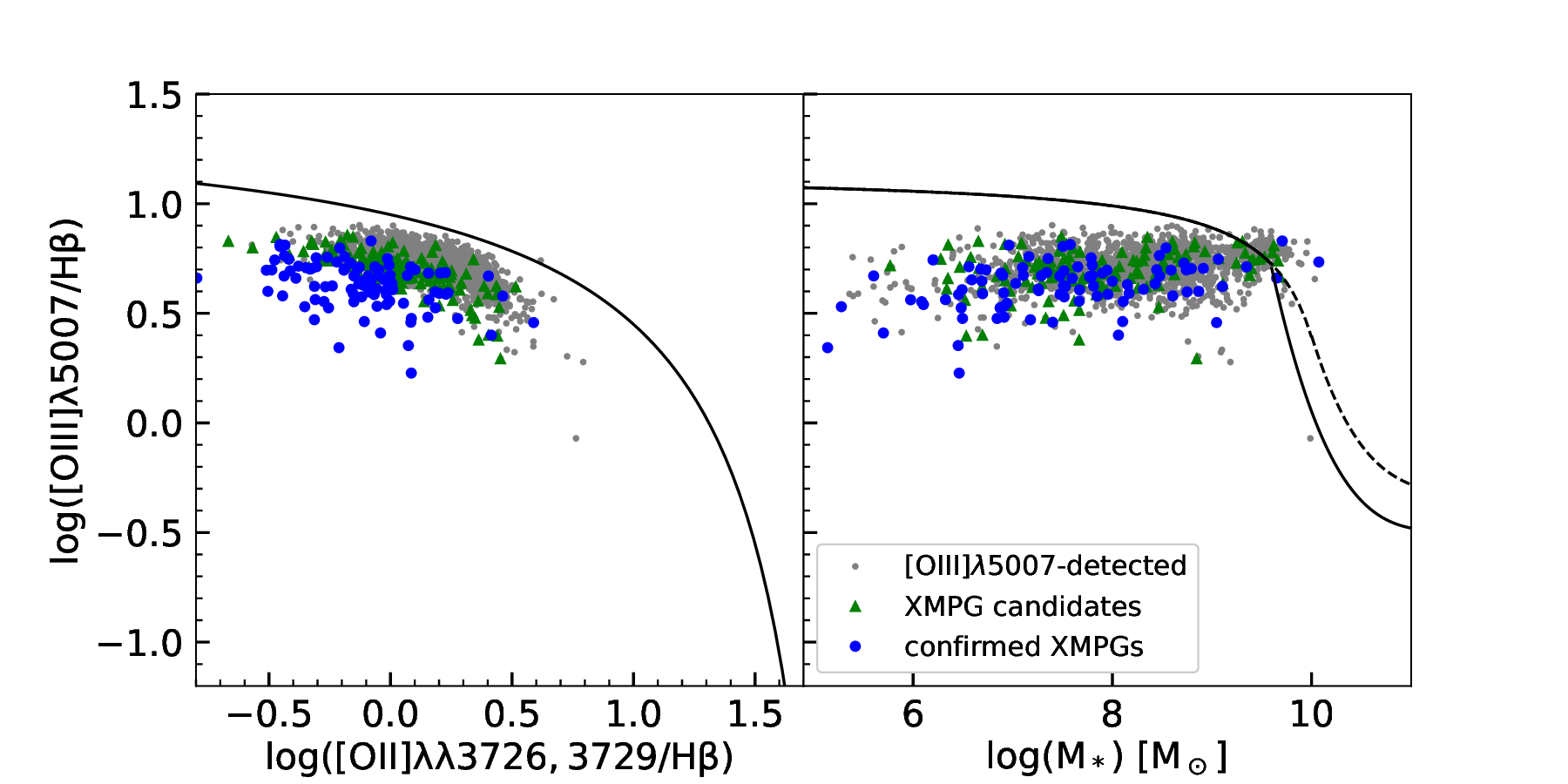}
\caption{Left: BPT diagnostic diagram of {\OIIIFIZ/\Hb} vs. {\OII/\Hb} for all \OIIIFOT-detected galaxies (grey), confirmed XMPGs (blue) and XMPG candidates (green). The solid curve is the demarcation line as defined in Equation (\ref{equ:bpt3}) to separate AGNs from star-forming galaxies. Right: MEx diagram for the same galaxies. The solid and dashed lines are the demarcation curves for starbursts \boldtext{(lower left) and AGNs (upper right)}, respectively, which come from \citet{Jun14}. \label{fig:bpt_xmpg}} 
\end{figure}

Figure \ref{fig:bpt_xmpg} displays the distribution of our XMPGs on the BPT diagram of {\OIIIFIZ/\Hb} vs. {\OII/\Hb}. Other \OIIIFOT-detected galaxies are also overplotted in this figure. Almost all \OIIIFOT-detected galaxies except one galaxy are star-forming galaxies lying below the demarcation line. This galaxy is selected as a star-forming galaxy via the BPT diagrams of \OIII/{\Hb} vs. \NII/{\Ha}  and \OIII/{\Hb} vs. \SII/\Ha. Compared to other \OIIIFIZ-detected galaxies, our XMPGs present slightly smaller \OIII/{\Hb}, but still have high ionization conditions due to large \OIII/{\OII} ratios. We also check the loci of these galaxies in the Mass-Excitation (MEx) diagnostic diagram \citep{Jun11,Jun14} and find that only 2.5\% of our XMPGs and 5.2\% of \OIIIFOT-detected galaxies might host an AGNs. Different diagnostic diagrams may have slightly different AGN predictions. In addition, there are parameter measurement uncertainties in these diagrams. So there could be some inconsistent classifications using different diagnostic diagrams. 

\subsection{Environment} \label{sec:env}
\begin{figure}[tbh!]
\centering
\includegraphics[width=1.0\columnwidth]{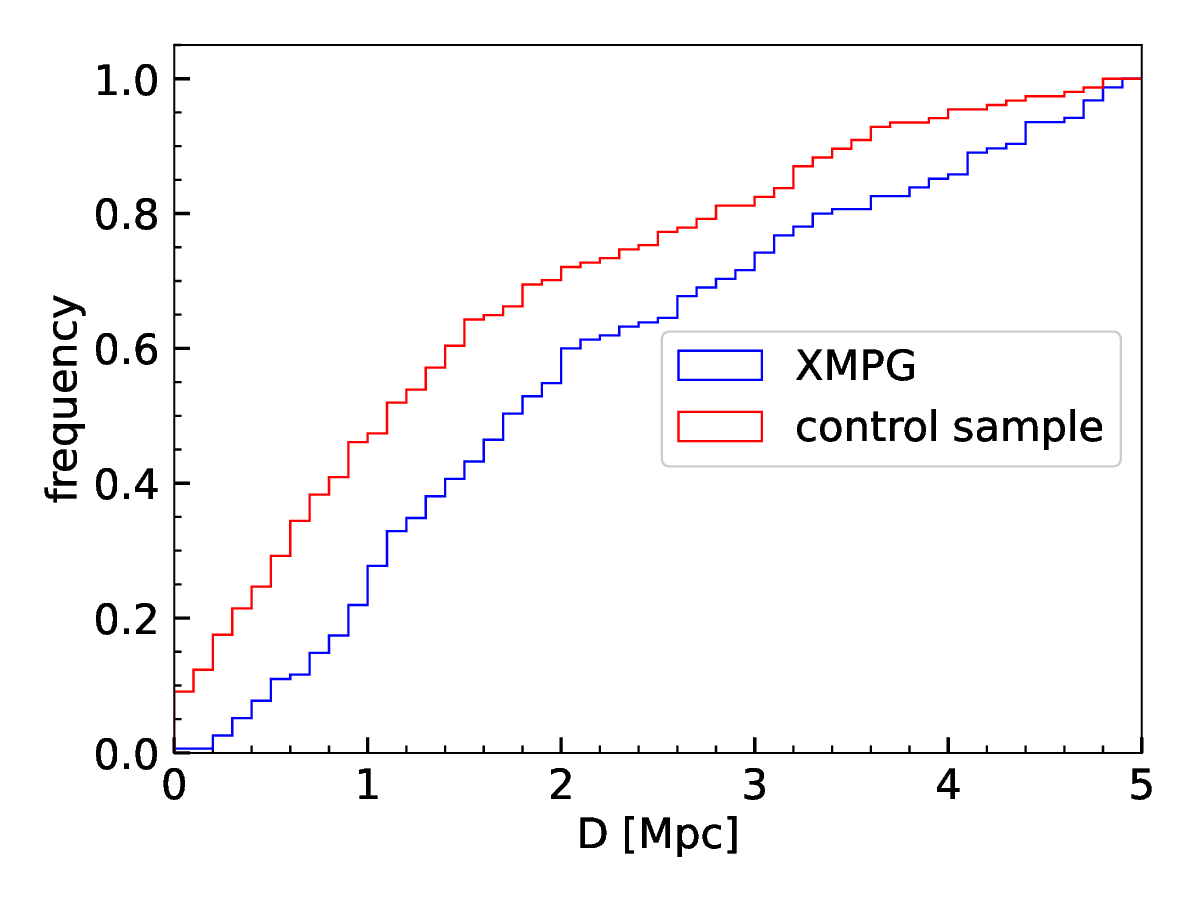}
\caption{Cumulative distributions of the closest distance ($D$) in all DESI star-forming galaxies to each galaxy in our XMPG (blue) and control (red) samples. \label{fig:env}}
\end{figure}

To characterize the environment of our XMPGs, we adopt the statistics adopted in the analyses of the so-called blueberry galaxies \citep{Yan17}, which is the distance ($D$) of the nearest galaxy. We construct a control sample from the DESI spectroscopic catalog for comparison. The comparison galaxy samples are selected through closest matching in both spatial distance and stellar mass with the total spectroscopically-confirmed star-forming galaxies. This makes sure that the control sample has both the same redshift and mass distributions as those of our XMPGs. Figure \ref{fig:env} shows the cumulative distributions of $D$ for these two samples. From this figure, we can see that there is an obvious difference between the two distributions. The Kolmogorov-Smirnov test gives a p-value close to zero, proving the significance. We conclude that our XMPGs inhabit relatively low-density environments. Nevertheless, we do not rule out that XMPGs might be mergers or closely interact with neighbour satellites, because the statistics used in this paper is on a large scale.

As investigated by \citet{Car09} and \citet{Yan17}, both green pea and blueberry galaxies are strong {\OIII} and compact emitters at different redshifts. These galaxies are relatively metal-poor and are found to be situated in low-density environments. More than half of our XMPGs have effective radii less than 3 kpc and equivalent width of {\OIIIFIZ} larger than 200 \AA. These features are used to select green pea-like galaxies. It could be foreseeable that our XMPGs are also located in low-density regions. In addition, \citet{Fil15} and \citet{San16a} explored the large-scale environment of XMPGs using the $N$-body cosmological numerical simulations. They found that XMPGs tend to reside in low-density environments similar to blue compact dwarf galaxies. About 75\% of their XMPGs have a strong tendency to reside in voids and sheets. 

\subsection{Mass-metallicity relation of XMPGs} \label{sec:relation}

It has been well established that the metallicity of star-forming galaxies correlates with the stellar mass \citep{Tre04}, forming the so-called mass-metallicity relation (MZR). The MZR reveals that more massive galaxies appear to be more chemically enriched and low-mass galaxies for a steeper relation in this mass-metallicity plane. It was also found that the MZR evolves with redshift. The relation still holds at high redshifts out to 3 and beyond, while high-redshift galaxies are observed to be less enriched \citep{Erb06,Mai08,Mou11,Zah13,LyC16}. Analytic models and numerical simulations of galactic chemical evolution have been developed to interpret the MZR \citep{Dav11,Lil13,Tor18}. These theoretical models disclose that the existence of MZR is the consequence of the secular interplay among different physical processes including stellar formation, gas outflow, and gas recycling and accretion. 

\begin{figure}[htb!]
\centering
\includegraphics[width=1.0\columnwidth]{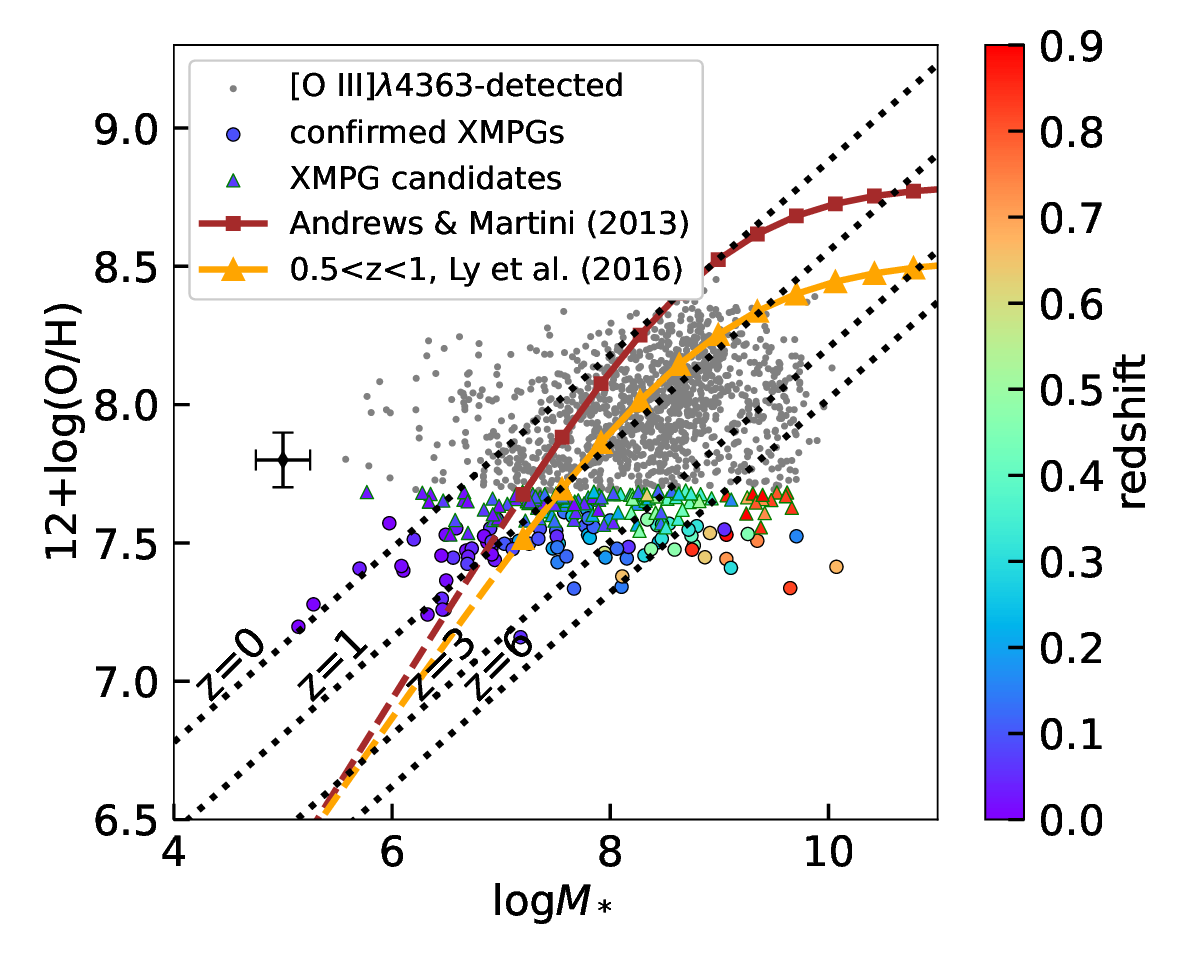}
\caption{Stellar mass-metallicity relation of our XMPGs color-coded by redshift. Smaller grey points are the {\OIIIFOT}-detected galaxies. The point with error bars shows the typical parameter measurement uncertainties. The local MZR from \citet{And13} at $z\sim0.08$ is plotted in brown with squares. The MZR at $0.5<z<1$ of \citet{LyC16} is plotted in orange with triangles. The dashed extended lines corresponds to extrapolation of these two relations to the low-mass region.  The dotted lines show the MZRs of theoretical simulations at $z=0,1,3,$ and 6 from \citet{MaX16}.  \label{fig:mzr}}
\end{figure}

Figure \ref{fig:mzr} presents the MZR of our XMPGs (colored points) and all \OIIIFOT-detected galaxies (grey points). It can be seen that our XMPGs at higher redshift tend to be the galaxies with higher stellar mass, mainly due to the observational selection effect. In much MZR research, the metallicity determinations are based on strong-line calibrations. However, such determinations generate nearly 1 dex differences depending on the choice of calibration method \citep{Kew08,Mou10}. For this reason, the MZRs based on the $T_e$ method are chosen for consistent comparisons. \citet{And13} (hereafter AM13) obtained a MZR at $z\sim0.08$ via the direct method for normal star-forming galaxies using stacked SDSS spectra. This MZR, spanning a wide stellar mass range of $\log{M_*} =$7.4--10.5, is plotted in Figure \ref{fig:mzr}. We also display the MZR at $0.5<z<1$ derived by \citet{LyC16}, who used the direct $T_e$ method to measure the metallicity for several tens of galaxies at $z < 1$ with Keck/MMT spectra. From this figure, we can see that \OIIIFOT-detected galaxies tend to lie below the local and \citet{LyC16} MZRs, as expected since the {\OIIIFOT} line is more easily detected in low-metallicity galaxies.

Our XMPGs cover a wide range of stellar masses, from $M_*=\sim 10^5$ to $\sim10^{10}\Msun$. About 44\% of our XMPGs at $z<0.25$ are located below the AM13 MZR, and 100\% of our XMPGs with $z>0.5$ are located below the \citet{LyC16} MZR, taking the MZR uncertainties (about 0.2 dex) into account. The deviations of our XMPGs from the MZRs of normal star-forming galaxies indicate that they might have distinct evolutionary pathways or reside in different environments. In addition, we also present the MZR evolution from the simulations of \citet{MaX16}. Their MZRs at $z=0,1,3$, and 6 are shown in Figure \ref{fig:mzr}. These simulations succeed in reproducing many observed properties (including MZR) of normal star forming galaxies at $0<z<6$. It can be clearly seen that a considerable number of our XMPGs are located within or out of the MZR at $z=$3--6. These XMPGs are probably the analogs of high-$z$ young galaxies, which are excellent samples for studying the evolution stage of the early universe.

In this analysis of the MZR relation, it is impossible to completely avoid sample selection effect. Our XMPG galaxy sample comes from the DESI early data, which includes the observations from the survey validation (SV) and first two months of the main surveys. The sample selection is extremely complicated because of different observing strategies during different SV stages and the main survey \citep{Des23a}, different selection methods of galaxy targets, the incompleteness from the fiber assignment and redshift measurement, and the bias induced by the XMPG detection method. We use the {\OIIIFOT} line to identify XMPGs in this paper. This emission line is very weak and tends to be detectable in metal-poor galaxies, so it bias the galaxies with low metallicity. Nevertheless, it is obvious that the XMPGs deviate from the normal MZR, illustrating their peculiarity. It is also possible that the lower MZR of our samples relative to normal galaxies is a physical trend. \citet{Jun14} mentioned that there might be a relation between the emission-line luminosity and the metallicity trends. The low-redshift galaxies with high emission-line luminosity (like our metal-poor galaxy samples) follow a systematically lower MZR, similar to those galaxies at high redshift. \boldtext{It has been also observed that the MZR has a secondary parameter dependency on the SFR \citep{Ell08, Man10}. There is a more fundamental mutual relationship between $M_*$, metallicity and SFR, that is so-called fundamental metallicity relation (FMR). It is found that there is no significant redshift evolution of FMR up to $z\sim3$ \citep{Man10}. The existence of FMR means that galaxies with higher sSFR galaxies present lower metallicities.} 

\section{Summary} \label{sec:summary}
In this work, we search for and identify a large sample of XMPGs at $z <1$ from the early data of the DESI project and investigate their stellar mass-metallicity relation. It is expected that several thousands of XMPGs will be discovered from the full DESI survey in the future. 

From the total number of 2.85 million galaxies in the DESI early data, we apply some SNR and FWHM cuts to select galaxies with good detections of emission lines and exclude galaxies with spurious line detection or broad-line AGNs. Diagnostic BPT diagrams are used to get rid of narrow-line AGNs. The final sample includes 1,623 star-forming galaxies with significant {\OIIIFOT} detections. We adopt the direct {\Te} method to measure the oxygen abundance, which relies on the detection of the {\OIIIFOT} auroral line. The stellar masses of the galaxies are estimated from broad-band photometry and the DESI spectra. 

A total of 223 XMPGs with oxygen abundance $Z < 0.1\Zsun$ are identified, including the largest sample of 58 XMPGs at $z>0.3$. If taking the metallicity uncertainty into account, we get 95 confirmed XMPGs. The remaining 128 galaxies are XMPG candidates regardless of the uncertainty. Most of our XMPGs are dust-free dwarfs and have large EW([OIII]$\lambda$5007) similar to green-pea galaxies with strong {\OIII} emissions. The most XMPG has a metallicity of about 1/34 of the solar value. A preliminary imaging examination of the two most metal-poor galaxies discovered in this study reveals two different morphologies, possibly indicating different evolution and physical origins. This will be investigated further in the future. The color-color diagram of $g-r$ vs. $r-z$ shows a wide color distribution of our XMPGs in addition to the low-redshift sequence of the literature XMPGs. This is helpful to select XMPG targets at relatively high redshift for future spectroscopic follow-up. We characterize the local environment of our XMPGs through calculating the distance to the closest galaxy. In contrast with the control sample of normal star-forming galaxies, it is confirmed that our XMPGs tend to reside in relatively low-density environments.

A significant number of our XMPGs are located far below the local MZR and the MZR of normal star-forming galaxies in the similar redshift range. Comparing with the simulations of \citet{MaX16}, our XMPGs are possibly low-redshift analogs of galaxies at high redshifts reaching $z=6$ or beyond. They are excellent objects for exploring the evolution stage of the early universe. Future detailed investigations of these samples will provide critical constraints on this galaxy population.  


\begin{acknowledgments}
This work is supported by the National Key R\&D Program of China (grant No. 2022YFA1602902), National Natural Science Foundation of China (NSFC; grant Nos. 12120101003, 12373010, and 11890691), and Beijing Municipal Natural Science Foundation (grant No. 1222028). We acknowledge the science research grants from the China Manned Space Project with Nos. CMS-CSST-2021-A02 and CMS-CSST-2021-A04. MS acknowledges the supports from the Polish National Agency for Academic Exchange (Bekker grant BPN/BEK/2021/1/00298/DEC/1), the European Union's Horizon 2020 Research and Innovation programme under the Maria Sklodowska-Curie grant agreement (No. 754510).

This research is supported by the Director, Office of Science, Office of High Energy Physics of the U.S. Department of Energy under Contract No. DE-AC02-05CH11231, and by the National Energy Research Scientific Computing Center, a DOE Office of Science User Facility under the same contract; additional support for DESI is provided by the U.S. National Science Foundation, Division of Astronomical Sciences under Contract No. AST-0950945 to the NSF’s National Optical-Infrared Astronomy Research Laboratory; the Science and Technologies Facilities Council of the United Kingdom; the Gordon and Betty Moore Foundation; the Heising-Simons Foundation; the French Alternative Energies and Atomic Energy Commission (CEA); the National Council of Science and Technology of Mexico (CONACYT); the Ministry of Science and Innovation of Spain (MICINN), and by the DESI Member Institutions: \url{https://www.desi.lbl.gov/collaborating-institutions}.

The DESI Legacy Imaging Surveys consist of three individual and complementary projects: the Dark Energy Camera Legacy Survey (DECaLS), the Beijing-Arizona Sky Survey (BASS), and the Mayall $z$-band Legacy Survey (MzLS). DECaLS, BASS and MzLS together include data obtained, respectively, at the Blanco telescope, Cerro Tololo Inter-American Observatory, NSF’s NOIRLab; the Bok telescope, Steward Observatory, University of Arizona; and the Mayall telescope, Kitt Peak National Observatory, NOIRLab. NOIRLab is operated by the Association of Universities for Research in Astronomy (AURA) under a cooperative agreement with the National Science Foundation. Pipeline processing and analyses of the data were supported by NOIRLab and the Lawrence Berkeley National Laboratory. Legacy Surveys also use data products from the Near-Earth Object Wide-field Infrared Survey Explorer (NEOWISE), a project of the Jet Propulsion Laboratory/California Institute of Technology, funded by the National Aeronautics and Space Administration. Legacy Surveys was supported by: the Director, Office of Science, Office of High Energy Physics of the U.S. Department of Energy; the National Energy Research Scientific Computing Center, a DOE Office of Science User Facility; the U.S. National Science Foundation, Division of Astronomical Sciences; the National Astronomical Observatories of China, the Chinese Academy of Sciences and the Chinese National Natural Science Foundation. LBNL is managed by the Regents of the University of California under contract to the U.S. Department of Energy. The complete acknowledgments can be found at \url{https://www.legacysurvey.org/}.

The authors are honored to be permitted to conduct scientific research on Iolkam Du’ag (Kitt Peak), a mountain with particular significance to the Tohono O’odham Nation.

\end{acknowledgments}

%






\appendix
\section{Basic information and properties for our confirmed XMPGs and XMPG candidates} \label{apx:catalog}
Table \ref{tab:xmpgs} and \ref{tab:xmpgs_can} list the basic information and measured properties for our 146 confirmed XMPGs and 200 XMPG candidates, respectively. The data in these two tables and all data points shown in the published graphs are available in a machine-readable form in Zenodo at \url{https://zenodo.org/record/8340116}.
\begin{longrotatetable}
\begin{deluxetable*}{crrccccccccccc}
\centering
\tablecaption{The properties of our confirmed XMPGs}
\label{tab:xmpgs}
\setlength\tabcolsep{2pt}
\tablewidth{0pt}
\tabletypesize{\tiny}
\tablehead{
\colhead{No} & \colhead{Name} & \colhead{TARGETID} & \colhead{RA} & \colhead{DEC} & \colhead{redshift} & \colhead{$r$} & \colhead{E(B-V)}  & \colhead{EW(\OIII)} & \colhead{$N_e$} & \colhead{$T_e(\OIII)$} & \colhead{12+log(O/H)} & \colhead{$M_*$} & \colhead{log(SFR)} \\
\colhead{} &  \colhead{} & \colhead{} &  \colhead{degree} &  \colhead{degree} &  \colhead{} &  \colhead{mag} &  \colhead{mag} &  \colhead{{\AA}} &  \colhead{cm$^{-3}$} &  \colhead{K} &  \colhead{} &  \colhead{log(\Msun)} &  \colhead{log({\Msun} yr$^{-1}$)} } 
\decimalcolnumbers
\startdata  
    1 &  DESIJ150535.89+314639.4 &  39628517750604871 & 226.39956 &  31.77763 & 0.0543 & 20.53 &            0.00 &     2.27 & 3.62e+02 $\pm$ 3.18e+02 & 2.33e+04 $\pm$ 1.30e+03 & 7.16 $\pm$ 0.05 &  7.18 $\pm$ 0.23 & -0.66 $\pm$  0.13 \\  
    2 &  DESIJ092331.28+645111.3 &  39633440282250119 & 140.88032 &  64.85315 & 0.0054 & 20.31 &            0.00 &     2.27 & 2.84e+02 $\pm$ 2.30e+02 & 2.07e+04 $\pm$ 1.89e+03 & 7.20 $\pm$ 0.09 &  5.14 $\pm$ 0.27 & -2.47 $\pm$  0.13 \\  
    3 &  DESIJ213658.82+041404.0 &  39627892803503091 & 324.24507 &   4.23445 & 0.0168 & 19.02 & 0.11 $\pm$ 0.03 &     2.93 & 4.92e+02 $\pm$ 4.18e+02 & 2.34e+04 $\pm$ 1.19e+03 & 7.24 $\pm$ 0.05 &  6.32 $\pm$ 0.09 & -0.61 $\pm$  0.13 \\  
    4 &  DESIJ130103.92+250558.2 &  39628373655292169 & 195.26632 &  25.09950 & 0.0257 & 20.68 & 0.00 $\pm$ 0.05 &     2.17 & 3.93e+02 $\pm$ 2.88e+02 & 2.02e+04 $\pm$ 3.15e+03 & 7.26 $\pm$ 0.17 &  6.46 $\pm$ 0.30 & -1.21 $\pm$  0.13 \\  
    5 &  DESIJ120751.02+002547.5 &  39627799857730585 & 181.96260 &   0.42989 & 0.0809 & 22.63 & 0.29 $\pm$ 0.19 &     2.86 &                     100 & 2.20e+04 $\pm$ 2.42e+03 & 7.26 $\pm$ 0.11 &  6.48 $\pm$ 0.23 & -0.36 $\pm$  0.13 \\  
    6 &  DESIJ144421.75+072050.7 &  39627963238450509 & 221.09063 &   7.34742 & 0.0058 & 19.39 &            0.00 &     2.76 &                     100 & 2.12e+04 $\pm$ 1.50e+03 & 7.28 $\pm$ 0.07 &  5.28 $\pm$ 0.11 & -1.83 $\pm$  0.13 \\  
    7 &  DESIJ125640.82+243304.2 &  39628362649440128 & 194.17009 &  24.55117 & 0.0488 & 21.80 & 0.02 $\pm$ 0.06 &     2.61 &                     100 & 2.33e+04 $\pm$ 1.56e+03 & 7.30 $\pm$ 0.06 &  6.46 $\pm$ 0.25 & -1.11 $\pm$  0.13 \\  
    8 &  DESIJ083738.65+314918.2 &  39628516362290842 & 129.41106 &  31.82174 & 0.3502 & 22.38 &            0.00 &     2.62 & 3.54e+02 $\pm$ 3.24e+02 & 2.13e+04 $\pm$ 2.74e+03 & 7.32 $\pm$ 0.11 &  8.61 $\pm$ 0.25 &  0.36 $\pm$  0.13 \\  
    9 &  DESIJ090136.94+333611.4 &  39632950190410845 & 135.40392 &  33.60317 & 0.1162 & 22.00 & 0.05 $\pm$ 0.14 &     2.32 &                     100 & 2.20e+04 $\pm$ 2.02e+03 & 7.34 $\pm$ 0.09 &  7.67 $\pm$ 0.22 & -0.46 $\pm$  0.13 \\  
   10 &  DESIJ085602.66+310445.1 &  39628500923056706 & 134.01108 &  31.07921 & 0.8221 & 22.90 & 0.02 $\pm$ 0.12 &     2.68 &                     100 & 2.22e+04 $\pm$ 1.32e+03 & 7.34 $\pm$ 0.05 &  9.65 $\pm$ 0.39 &  1.37 $\pm$  0.13 \\  
\enddata
\tablecomments{(1) Sequence number; (2) defined XMPG name; (3) DESI target ID; (4) and (5) respectively for right ascension and declination in J2000; (6) spectroscopic redshift; (7) $r$-band magnitude; (8) gas-phase extinction; (9) equivalent widths of \OIII$\lambda$5007; (10) electronic density; (11) electronic temperature for {\OIII}; (12) oxygen abundance; (13) stellar mass; (14) logarithmic SFR;  Only top 10 of the whole XMPGs sorted by increasing 12+log(O/H) are listed. \\
(This table is available in its entirety in machine-readable form.)}
\end{deluxetable*}
\end{longrotatetable}


\begin{longrotatetable}
\begin{deluxetable*}{crrccccccccccc}
\centering
\tablecaption{The properties of our XMPG candidates}
\label{tab:xmpgs_can}
\setlength\tabcolsep{2pt}
\tablewidth{0pt}
\tabletypesize{\tiny}
\tablehead{
\colhead{No} & \colhead{Name} & \colhead{TARGETID} & \colhead{RA} & \colhead{DEC} & \colhead{redshift} & \colhead{$r$} & \colhead{E(B-V)}  & \colhead{EW(\OIII)} & \colhead{$N_e$} & \colhead{$T_e(\OIII)$} & \colhead{12+log(O/H)} & \colhead{$M_*$} & \colhead{log(SFR)} \\
\colhead{} &  \colhead{} & \colhead{} &  \colhead{degree} &  \colhead{degree} &  \colhead{} &  \colhead{mag} &  \colhead{mag} &  \colhead{{\AA}} &  \colhead{cm$^{-3}$} &  \colhead{K} &  \colhead{} &  \colhead{log(\Msun)} &  \colhead{log({\Msun} yr$^{-1}$)} } 
\decimalcolnumbers
\startdata  
    1 &  DESIJ140344.61+484557.5 &  39633228671224911 & 210.93585 &  48.76599 & 0.0025 & 16.87 &            0.00 &     2.06 & 3.19e+02 $\pm$ 2.49e+02 & 1.96e+04 $\pm$ 3.27e+03 & 7.53 $\pm$ 0.17 &  6.53 $\pm$ 0.23 & -2.15 $\pm$  0.13 \\ 
    2 &  DESIJ072211.92+415915.2 &  39633112598056268 & 110.54969 &  41.98758 & 0.0438 & 20.65 &            0.00 &     2.33 & 1.95e+02 $\pm$ 1.31e+02 & 1.88e+04 $\pm$ 3.40e+03 & 7.54 $\pm$ 0.19 &  6.70 $\pm$ 0.23 & -0.82 $\pm$  0.13 \\ 
    3 &  DESIJ140237.80+354229.6 &  39632996071902273 & 210.65748 &  35.70824 & 0.1170 & 24.93 &            0.00 &     2.75 &                     100 & 1.96e+04 $\pm$ 3.32e+03 & 7.54 $\pm$ 0.16 &  6.76 $\pm$ 0.51 & -1.46 $\pm$  0.13 \\ 
    4 &  DESIJ143510.37+030408.5 &  39627860876461065 & 218.79323 &   3.06904 & 0.4578 & 23.01 & 0.01 $\pm$ 0.18 &     2.89 & 6.28e+02 $\pm$ 5.02e+02 & 1.92e+04 $\pm$ 3.06e+03 & 7.54 $\pm$ 0.15 &  8.27 $\pm$ 0.33 &  0.67 $\pm$  0.13 \\ 
    5 &  DESIJ140128.57+041520.0 &  39627890895095223 & 210.36905 &   4.25556 & 0.1170 & 23.33 &            0.00 &     2.22 & 2.90e+02 $\pm$ 2.06e+02 & 1.86e+04 $\pm$ 2.74e+03 & 7.55 $\pm$ 0.16 &  7.36 $\pm$ 0.27 & -1.14 $\pm$  0.13 \\ 
    6 &  DESIJ084809.70+313439.3 &  39628511245239168 & 132.04043 &  31.57760 & 0.8516 & 23.49 & 0.11 $\pm$ 0.19 &     3.00 & 6.30e+02 $\pm$ 4.87e+02 & 1.89e+04 $\pm$ 2.45e+03 & 7.56 $\pm$ 0.14 &  9.38 $\pm$ 0.33 &  1.37 $\pm$  0.13 \\ 
    7 &  DESIJ081428.78+342638.3 &  39632970113354267 & 123.61991 &  34.44398 & 0.3236 & 22.32 &            0.00 &     2.85 & 8.51e+02 $\pm$ 7.19e+02 & 2.00e+04 $\pm$ 3.08e+03 & 7.56 $\pm$ 0.15 &  8.51 $\pm$ 0.24 &  0.28 $\pm$  0.13 \\ 
    8 &  DESIJ160521.36+422914.4 &  39633123201254715 & 241.33899 &  42.48735 & 0.0775 & 18.71 & 0.10 $\pm$ 0.05 &     1.92 & 7.56e+01 $\pm$ 7.58e+01 & 1.84e+04 $\pm$ 3.42e+03 & 7.56 $\pm$ 0.22 &  8.85 $\pm$ 0.18 &  0.30 $\pm$  0.13 \\ 
    9 &  DESIJ144707.28+062409.9 &  39627945244886188 & 221.78032 &   6.40276 & 0.0726 & 20.12 & 0.01 $\pm$ 0.06 &     2.25 &                     100 & 1.94e+04 $\pm$ 2.95e+03 & 7.57 $\pm$ 0.15 &  7.94 $\pm$ 0.22 & -0.36 $\pm$  0.13 \\ 
   10 &  DESIJ122340.63-060720.9 &  39627643062062197 & 185.91928 &  -6.12249 & 0.2067 & 23.43 &            0.00 &     2.32 & 2.77e+02 $\pm$ 2.61e+02 & 2.07e+04 $\pm$ 2.81e+03 & 7.57 $\pm$ 0.14 &  8.25 $\pm$ 0.21 & -0.85 $\pm$  0.13 \\ 
\enddata
\tablecomments{Only top 10 of the whole XMPGs sorted by increasing 12+log(O/H) are listed. \\
(This table is available in its entirety in machine-readable form.)}
\end{deluxetable*}
\end{longrotatetable}

\bibliography{xmpg_sample}{}
\bibliographystyle{aasjournal}



\end{document}